\documentclass[12pt,preprint]{aastex}

\newcommand{\be}{\begin{equation}}
\newcommand{\ee}{\end{equation}}
\newcommand{\fone}{ {\cal F}_1 } 
\newcommand{\cross}{{ \langle \sigma \rangle_{ss} }} 
\newcommand{\exgamma}{{ \eta }}
\newcommand{\nchar}{{ \langle n_\ast \rangle }}

\begin{document}

\title{The Birth Environment of the Solar System} 

\author{Fred C. Adams} 
 
\affil{Michigan Center for Theoretical Physics \\  
Physics Department, University of Michigan, Ann Arbor, MI 48109, USA}  

\affil{to appear in Annual Reviews of Astronomy and Astrophysics (2010, Volume 48)} 

\begin{abstract}

This paper reviews our current understanding of the possible birth
environments of our Solar System. Since most stars form within groups
and clusters, the question becomes one of determining the nature of
the birth aggregate of the Sun. This discussion starts by reviewing
Solar System properties that provide constraints on our environmental
history. We then outline the range of star-forming environments that
are available in the Galaxy, and discuss how they affect star and
planet formation.  The nature of the solar birth cluster is
constrained by many physical considerations, including radiation
fields provided by the background environment, dynamical scattering
interactions, and by the necessity of producing the short-lived
radioactive nuclear species inferred from meteoritic measurements.
Working scenarios for the solar birth aggregate can be constructed, 
as discussed herein, although significant uncertainties remain. 

\end{abstract}

\keywords{Cluster Dynamics, Nuclear Abundances, Planet Formation, 
Star Formation, Stellar Clusters, Supernovae, and the Sun}  

\section{INTRODUCTION}  

Some of the most foundational questions in astrophysics are those of
``origins'', including the formation of the universe, galaxies, stars,
and planets. On each of these fundamental scales, astronomical
entities are brought into existence through complex physical
processes, live out their lives, and often end with death-like
finality. The origin of the universe and galaxy formation fall in the
domain of cosmology, although these scales are largely decoupled from
the question of solar birth. On smaller scales, star formation and
planet formation are intimately connected, since planets form within
the circumstellar disks that arise from protostellar collapse. In this
regime, the origin of our own star and its planetary system represents
a fundamental astronomical issue.

Recent studies have underscored the finding that most stars are not
born in isolation, but rather form within groups and clusters (e.g.,
Carpenter 2000, Lada \& Lada 2003, Porras et al. 2003). Although some
fraction of the stellar population does form in an isolated mode
(e.g., in the nearby Taurus star-forming region), the considerations
of this review suggest that our Solar System formed within a group or
cluster with some membership size $N$. This state of affairs thus
poses a number of coupled questions considered herein: What was the
size $N$ of the solar birth cluster? What constraints can we place on
the properties of the solar birth aggregate? How did the solar birth
environment influence the formation of our planetary system? How rare
or common are the necessary conditions that gave rise to our Solar
System?

At the current epoch, the Solar System is nearly five billion years
old (more precisely, about 4600 Myr).  As outlined above, stars form
in groups and clusters that are embedded within molecular clouds, and
these clouds have lifetimes measured in only tens of millions of years
(or even less --- see Hartmann et al. 2001).  Within the clouds, the
clusters themselves also live for tens of millions of years, or
somewhat shorter times. Even the long-lived open clusters dynamically
evaporate over hundreds of millions of years. As a result, the birth
environment of the Sun has long since been dissipated.  Nonetheless,
as reviewed herein, the extant properties of our Solar System, coupled
with our emerging understanding of star and planet formation
processes, allow us to reconstruct some of the requirements of the
birthplace of the Sun.

One reason for constraining the solar birth environment is because of
its intrinsic interest.  In addition to the issue of where we came
from, however, the origin of our Solar System provides an important
consistency check on the current paradigms of star and planet
formation. These theories have been tested and modified over the past
decades, and now provide a compelling picture for the origin of stars
and planets. However, the properties of our Solar System are known in
much greater detail than those in other systems. As a result, our
Solar System represents an important additional test of the
theoretical framework. In particular, we would like our theories of
star/planet formation to produce solar systems with roughly the
properties of our own under unremarkable circumstances, i.e., under
conditions that are relatively common in observed star forming
regions.

\subsection{Scope of this Review} 

This review focuses on three general types of physical process that
influence solar system formation and thereby constrain the possible
birth environment of the Sun:

[1] The birth cluster affects star and planet formation through
dynamical processes, including disk truncation and perturbations of
planetary orbits due to passing stars. The observed lack of severe
disruption provides an upper limit on the density of the birth cluster
and the time that the Solar System lived in such an environment. On
the other hand, the observed orbital elements of Sedna (and perhaps
other Kuiper Belt objects) can be explained by such close encounters,
so that some type of dynamical interactions may be required.

[2] The birth cluster also provides strong background radiation fields,
including those at ultraviolet (UV) wavelengths.  These fields can
drive the evaporation of the early solar nebula and hence the loss of
planet forming potential. Gas removal from the outer nebula also
affects the edge structure of the Solar System.  Harder radiation
(X-rays and EUV photons) provides ionization, which can influence both
the early star formation process and the subsequent evolution of our
circumstellar disk.

[3] The presence of short-lived radioactive species inferred from
meteorites provides another class of constraints. Radioactive nuclei
can be produced in supernova explosions, evolved stars, and by
internal irradiation mechanisms. Under the assumption that some
external enrichment is necessary, we obtain further constraints. In
particular, the cluster must be relatively large, the Sun must reside
within a confined range of radial locations, and the proper timing of
events must be realized.

This review considers these three classes of constraints, and includes
a general assessment of how star forming clusters can potentially
affect their constituent solar systems. In this respect, this review
differs from many previous discussions that primarily focus on the
origin of the short-lived radio isotopes (Wadhwa et al. 2007; Goswami
\& Vanhala 2000; Busso et al. 1999, 2003; Wasserburg et al. 2006)
and/or the thermodynamic history of the early solar system (e.g., Krot
et al. 2005ab, 2008). Although these issues are discussed herein, the 
reader is referred to these earlier reviews for additional details
regarding radio isotopes and heating in the early solar nebula (see
also Montmerle et al. 2006).

This discussion of the solar birth aggregate involves two coupled
themes: First, significant tension exists between the apparent need
for a large birth cluster to provide nuclear enrichment and to explain
Sedna's orbit, and the need for a smaller, less interactive cluster to
avoid overly disruptive dynamical and radiative effects. The required
compromise underscores the need for quantitative assessments of the
physical processes that inform the properties of the solar birth
environment. Second, as outlined below, many of these constraints must
be addressed statistically. For example, clusters do not fully sample
the high end of the stellar IMF, so that supernova enrichment does not
occur with certainty, but rather with a well-defined probability
distribution. Similarly, clusters are highly chaotic systems, so that
dynamical issues must also be addressed in terms of probabilities.

\subsection{Overview} 

This paper is organized as follows. In Section 2, we outline some of
the basic characteristics of our Solar System, with a focus on those
properties that provide clues to the past. The range of possible birth
environments is discussed in Section 3, along with an overview of
cluster properties.  The constraints on these clusters provided by
dynamical considerations are discussed in Section 4.  The radiation
fields produced by various star forming environments are then
considered in Section 5, along with an overview of their possible
effects on young and forming stars.  These results are then combined
with Solar System properties to obtain further constraints on the
birth environment. The question of short-lived radioactive isotopes,
as inferred from meteoritic data, is taken up in Section 6. This
discussion includes the two leading explanations for these
radionuclides, internal irradiation mechanisms and external enrichment
through supernovae. The review concludes in Section 7 with an overview
of the constraints, possible scenarios for the solar birth aggregate,
and a discussion of the general implications for star/planet
formation.
 
\section{SOLAR SYSTEM PROPERTIES} 

\subsection{The Sun and the IMF} 

The Sun is a relatively large star. Of the 50 nearest stars, the Sun
ranks a respectable fourth largest in terms of mass (e.g., see Henry
et al. 1994 and subsequent papers).  In contrast, the distribution of
stellar masses at the epoch of formation, the stellar IMF, is heavily
weighted toward stars of low mass (e.g., Scalo 1998).  A useful way to
parameterize the stellar IMF is to define $\fone$ to be the fraction
of the stellar population with masses greater than the Sun.
Observations indicate that $\fone \approx 0.12$. In other words, the
fact that our Sun is a relatively large star is a $\sim$12 percent
effect. As a result, the mass of the Sun is somewhat large but still
unremarkable.

In considering the possible range of birth environments, we need to
consider effects from the full distribution of stellar masses,
especially the high mass end. The largest stars provide the largest
potential impacts on star and planet formation, where these effects
include radiation, winds, and supernova explosions.  The mass
distribution for massive stars can be written in power-law form,
\be
{dN_\ast \over dm} = \fone \gamma m^{-(\gamma+1)} \, , 
\label{imf} 
\ee
where $m$ is the mass in Solar units and the index $\gamma \sim$ 1.35
(Salpeter 1955). Although the low-mass end of the IMF has received 
considerable attention in recent years, the power-law form for the
high-mass end remains robust. However, the value of the index $\gamma$
appears to have considerable scatter from region to region (Scalo 1998),
such that $\gamma$ is evenly distributed within the range $\gamma =
1.5 \pm 0.5$. For the sake of definiteness, in this review we use
$\gamma$ = 1.5 to characterize the high-mass end of the IMF, but the
reader should note that a range of values is allowed.

Note that the probability distribution of equation (\ref{imf}) is
normalized so that
\be
\int_1^\infty {dN_\ast \over dm} dm = \fone \, .
\ee 
Since the IMF has an upper mass limit at $m \equiv m_\infty \approx
100$, the above integral introduces a correction factor
$[1-(1/m_\infty)^{\gamma}] \approx 0.998$ in the normalization; we
neglect this correction in the present discussion.

\subsection{The Sun as a Single Star} 

The Sun is a single star, whereas the majority of solar-type stars are
found in binary systems (Abt 1983, Duquennoy \& Mayor 1991).  However,
about one third of solar-type stars are single, so that the lack of a
binary companion in our Solar System does not represent a significant
constraint. A single Sun corresponds to a $\sim$30 percent effect.
For completeness, we also note that since most stars are much smaller
than the Sun, and smaller stars are primarily single, the majority of
all stars are actually single (Lada 2006). Although the Sun
ended up as a single star, it remains possible for the Sun to have had
binary companions earlier in its history (see the discussion of
Malmberg et al. 2007). Any such companions must have had wide orbits,
however, so as not to disrupt the early solar nebula and/or the
planetary orbits. 

\subsection{Solar Metallicity} 

The Sun has relatively high metallicity (Wielen et al. 1996, Wielen \&
Wilson 1997). For G dwarfs in the Solar neighborhood, the metallicity
distribution has a peak at [Fe/H] = $-0.20$ dex (Rocha-Pinto \& Maciel 1996). 
For the same distribution, only about one fourth of the G dwarfs have 
metallicity as large as the Sun, so that the moderately high metal
content of the Solar System corresponds to a 25 percent effect. We
also note that some of the metallicity could be contributed by the
supernova that is thought to have enriched the solar system in
short-lived radioactive species. On a related note, the abundances of
oxygen isotopes in the Solar System, in particular the ratio
$^{18}$O/$^{17}$O, show interesting anomalies that could indicate
pollution by a nearby supernova (Young et al. 2009). On the other
hand, these anomalies could also be explained by isotope selective
photodissociation in the early solar nebula (Smith et al. 2009).

\subsection{Planets and their Orbits} 

One important feature of our Solar System is that it has produced a
substantial number of planets, including giant gaseous planets in the
outer regions and rocky terrestrial planets further inward. Although
observations of extra-solar planetary systems indicate that giant
planets are not rare, planet formation is not guaranteed. Current data
suggest that about 10 percent of solar-type stars harbor giant planets
with semi-major axes in the range $a$ = 0.02 -- 5 AU (Cummings et al.
2008).  Since the observational sample is not complete, especially for
planets with longer periods, the fraction of solar-type stars with
giant planets is even larger. After extrapolating the data to include
the full range of periods, the fraction of systems with giant planets
is estimated to be about 20 percent (although larger fractions,
perhaps up to 50 percent, remain possible).

At this time, detection of planets with masses comparable to Earth 
is just out of reach for main-sequence stars (due to technical 
limitations).  As a result, it is too early to assess the odds of
solar systems having terrestrial planets.  However, a number of
considerations suggest that such planets can be made relatively
easily: We first note that terrestrial mass planets have been detected
in orbit about pulsars (Wolszczan 1994).  In addition, our Solar
System has readily produced not only four terrestrial planets, but a
large collection of moons, asteroids, dwarf planets, and comets. Data
from extrasolar planets show that the planetary mass function
increases toward lower masses. These findings argue that the formation
of terrestrial planets should be common (and hence that our Solar
System is not rare in this regard).

One of the most remarkable features of our Solar System is that the
planetary orbits are well-ordered. More specifically, all of the
planetary orbits are nearly in the same plane, with inclination angles
in the narrow range $(\Delta i) \le 3.5^\circ$.  Another measure of
order is the low orbital eccentricities, which lie in the range 
$0 \le e \le 0.2$. If Mercury is excluded, the next largest
eccentricity is that of Mars at $e \approx 0.09$. As discussed below,
orbital eccentricities and inclination angles are relatively easy to
excite by passing stars and other dynamical perturbations. As a
result, the currently observed order of the Solar System provides
powerful constraints on its history.

On a related note, observations of extrasolar planets (e.g., Schneider
2009) show that planetary orbits in other solar systems display a wide
range of eccentricities. More specifically, for extrasolar planets
with semi-major axes $a \ge 0.1$ AU, the mean of the distribution is
approximately $\langle e \rangle \approx 0.3$ and the median is 
$e_h \approx 0.24$.  Orbits with smaller semi-major axes tend to be
circularized by tidal interactions with the central star.  The
inclination angles of extrasolar planetary systems are difficult to
measure. Nonetheless, available data indicate that the currently
observed multi-planet systems are more dynamically active, and
significantly less well-ordered, than our own (e.g., Udry \& Santos
2007).

\subsection{Edges of the Solar System} 

The Solar System has a number of outer ``edges'', which provide
further constraints on its dynamical past. The first obvious edge is
marked by the planet Neptune, which orbits with a semi-major axis 
$a \approx 30$ AU. The solar nebula must have extended out to 
(approximately) this radius in order to facilitate planet formation.

Beyond the last giant planet, the Solar System contains a large
collection of smaller rocky bodies in the Kuiper Belt. The orbits of
these Kuiper Belt objects, as a group, have much larger eccentricities
and inclination angles than the planetary orbits (Luu \& Jewitt 2002).
The Kuiper belt is thus dynamically excited, to a moderate degree, and
this property must be consistent with scenarios for solar birth.  In
spite of their large numbers, these bodies contain relatively little
total mass, which has been estimated to be 10 -- 100 times less than
the mass of Earth (Bernstein et al. 2004). The outer boundary of the
planetary system (and inner boundary of the Kuiper belt) at 30 AU is
thus significant. In addition, although the outer edge of this belt is
not perfectly sharp, a significant drop-off is observed around 50 AU
(Allen et al. 2000). This radial location roughly corresponds to the 
2:1 mean motion resonance with Neptune. Since, in principle, additional 
bodies could have formed and survived beyond this radius, the origin
of this edge at $\sim50$ AU represents an important issue.

At still greater distances, the Solar System contains a large,
nearly spherical collection of comets known as the Oort Cloud. This
structure extends to about 0.3 pc ($\sim$60,000 AU). Since the comets
in the Oort Cloud are loosely bound to the Sun, gravitational
perturbations from passing stars can easily disrupt the cloud.  From
the impulse approximation, the change in velocity of the Sun due to a
passing star is approximately given by $\Delta v$ = $2GM_\ast/(b
v_\infty)$, where $b$ is the distance of closest approach. Setting the
orbit speed (at radius $b$) of the comets equal to this change in
velocity, we obtain an estimate for the radius of the sphere in which
the Sun can protect its comets: 
\be
r \approx 
{4 G M_\ast^2 \over M_\odot v_\infty^2} \approx 0.017 {\rm pc} 
\left( {v_\infty \over 1 \, {\rm km} \, {\rm s}^{-1} } \right)^{-2} \, , 
\ee
where the numerical value assumes both the Sun and the passing star
have the same mass. For typical interaction rates, stellar densities,
and encounter speeds in embedded clusters, where the Sun might have
formed, most of the Oort cloud would be stripped by passing stars. 
As a result, this structure probably formed later, after the Sun left  
its birth environment.  In addition, the Oort cloud is likely to have
grown slowly, over perhaps 1 Gyr (Duncan et al. 1987), i.e., a timescale much
longer than the expected lifetime of the birth cluster.  Nonetheless,
the possible timing of close encounters is constrained by models of
Oort cloud formation (Levison et al. 2004, Brasser et al. 2006, Kaib
\& Quinn 2008).

\subsection{Minimum Mass Solar Nebula} 

Given the basic architecture of the Solar System described above, it
is customary to define a benchmark disk model known as the Minimum
Mass Solar Nebula (often denoted as MMSN). Since the planets are
currently enriched in heavy elements relative to the Sun, one must add
back in the mass of the gas required for the augmented system to have
solar metallicity. Although this exercise is not without
uncertainties, a standard version of the early solar nebula can be
defined. This disk has surface density $\Sigma (r)$ and temperature
$T(r)$ profiles of the power-law form 
\be
\Sigma(r) = \Sigma_1 \left( {1 \, {\rm AU} \over r} \right)^p 
\qquad {\rm and} \qquad 
T(r) = T_1 \left( {1 \, {\rm AU} \over r} \right)^q \, , 
\ee
where $\Sigma_1$ and $T_1$ are the values at $r$ = 1 AU, and where $p
\approx 3/2$ and $q \approx 1/2$. The surface density at 1 AU is
estimated to lie in the range $\Sigma_1 \approx 2000 - 4500$ g
cm$^{-2}$ (e.g., Weidenschilling 1977, Hayashi 1981). Using the upper
end of this range and assuming that the nebula extends out to $r_d$ =
30 AU, the enclosed mass is estimated to be $M_d \approx$ 0.035
$M_\odot$. Keep in mind that this model is only a reference point; for
example, recent observations suggest that $p \approx 1$ with lower
disk masses (Andrews et al. 2009), whereas disk accretion rates argue
for higher starting masses $M_d \approx 0.1 M_\odot$ (Hartmann
2007). Since planets can migrate, the starting surface density of the
nebula is subject to further uncertainty (compare Desch 2007 and Crida
2009).

\subsection{Short-lived Radioactive Isotopes}

One of the most intriguing --- and potentially constraining ---
properties of our Solar System is the inferred presence of short-lived
radioactive species during the epoch of planet formation.  For the
sake of definiteness, we consider ``short-lived'' species to be those
with half-lives less than about 10 Myr.  The presence of these radio
isotopes is inferred by measuring the daughter products in meteoritic
samples, which condensed into rocks during the formative stages of the
Solar System.  These short-lived radioactive species thus indicate
that only a short time (perhaps $\sim1$ Myr) could have elapsed between 
their production and their subsequent incorporation into early Solar
System material. A viable source of these short-lived radio isotopes
is thus indicated (for recent reviews, see Wasserburg et al. 2006,
Wadhwa et al. 2007, and references therein).

Table I lists some of the most important radio isotopes, including
$^{10}$Be, $^{26}$Al, $^{41}$Ca, $^{60}$Fe, $^{53}$Mn, $^{107}$Pd, and
$^{182}$Hf. The half-lives, daughter products, reference isotopes, and
fractional abundances are also given (see Wasserburg 1985, Cameron
1993, Goswami \& Vanhala 2000, McKeegan et al. 2000, and many others).
Additional measurements indicate the presence of radionuclides with
somewhat longer lifetimes, e.g., $^{129}$I, and $^{244}$Pu, with
half-lives of about 16 Myr and 80 Myr, respectively. Here we assume
that these longer-lived species can be explained by galactic-scale
nucleosynthesis coupled with mixing in the interstellar medium, so
that they do not constrain the birth environment (see Wasserburg et
al. 2006). In contrast, the short-lived species must be produced
locally, near the time and location of solar birth, i.e., on time
scales measured in Myr and distances measured in pc (see below). These
short-lived nuclei thus provide potential constraints on the birth
cluster.

\noindent
$\,$

\centerline{\bf Table I: Radio Isotopes} 

\bigskip
\begin{center} 
\begin{tabular}{ccccc} 
\hline
\hline 
Nuclear Species & Daughter & Reference & Half-life (Myr) & Mass Fraction \\ 
\hline
\hline
$^{7}$Be & $^{7}$Li & $^{9}$Be & 53 days & ($8 \times 10^{-13}$)\\ 
$^{10}$Be & $^{10}$B & $^{9}$Be & 1.5 & ($\sim10^{-13}$)\\ 
$^{26}$Al & $^{26}$Mg & $^{27}$Al & 0.72 & $3.8 \times 10^{-9}$ \\  
$^{36}$Cl & $^{36}$Ar & $^{35}$Cl & 0.30 & $8.8 \times 10^{-10}$ \\  
$^{41}$Ca & $^{41}$K & $^{40}$Ca & 0.10 & $1.1 \times 10^{-12}$ \\  
$^{53}$Mn & $^{53}$Cr & $^{55}$Mn & 3.7  & $4.0 \times 10^{-10}$ \\  
$^{60}$Fe & $^{60}$Ni & $^{56}$Fe & 1.5  & $1.1 \times 10^{-9}$ \\  
$^{107}$Pd & $^{107}$Ag & $^{108}$Pd & 6.5 & $9.0 \times 10^{-14}$ \\  
$^{182}$Hf & $^{182}$W & $^{180}$Hf & 8.9 & $1.0 \times 10^{-13}$ \\  
\hline  
\hline 
\end{tabular} 
\end{center}

\section{DEMOGRAPHICS OF STAR-FORMING REGIONS} 

\subsection{Distribution of Clusters} 

Given that most stars form within stellar groups and clusters, it is
useful to assess the possible range of these birth environments.
Unfortunately, at this time, the field has not reached a consensus
regarding the distribution of group/cluster sizes. The membership of
these aggregates varies from $N$ = 1 (corresponding to stars forming
in isolation) up to about $N \approx 10^6$ (corresponding to
proto-globular clusters). The relative frequency of these environments
is determined by the distribution $df_c/dN$, which is defined here to
be the probability that a cluster has membership size $N$. The
corresponding probability that a given star (or solar system) finds
itself born within a system of membership size $N$ is then given by 
\be
{dP \over dN} = N {df_c \over dN} \, , 
\ee 
with a normalization such that 
\be
\int_1^\infty {dP \over dN} dN = 1 \, . 
\ee
For surveys in the solar neighborhood, where clusters are found with
membership size in the range $N$ = 100 -- 2000, the distribution of
cluster number $df_c/dN \propto 1/N^2$, so that the probability
density for a star being born in a system of size $N$ has the form
$dP/dN \propto 1/N$ (e.g., see Carpenter 2000, Kroupa \& Boily 2002,
Lada \& Lada 2003, Porras et al. 2003).  The integral of this
distribution indicates that the cumulative probability $P \sim \ln N$, 
so that the probability of stars being born in birth aggregates of
varying size $N$ is evenly logarithmically distributed across the
range in $N$.

A similar result is found for larger clusters, where the observational
samples are farther away and less complete (especially at the lower
end of the distribution).  In this regime, observations suggest that
the distribution also has the form $df_c/dN \propto N^{-2}$ for values
of $N$ up to and including those of globular clusters (Elmegreen \& Efremov 1997).  
Given that both the solar neighborhood and the realm of large clusters
(roughly $N \ge 2000$) have distributions of the form $df_c/dN \sim
N^{-2}$, the simplest hypothesis is to assume that the same law holds
over the entire range of $N$. In this case, stars would form with
equal probability in each decade of $N$. This hypothesis is equivalent
to assuming that the normalizations of the distributions match up. For
purposes of estimating the probability of the Solar System being born
in clusters of various sizes, we will use this simple, log-random
distribution.  However, the reader should keep in mind that not enough
data exist for this hypothesis to be verified.

Two additional issues introduce further uncertainty in defining the
distribution of cluster sizes: 

[1] Although many discussions dismiss the importance of isolated star
formation, sometimes called the distributed population, results from
the {\it Spitzer} Space Telescope (which provides an unbiased survey)
indicate otherwise: Let $N_M$ be the ``median point'' of the
distribution, so that half of the stars form within clusters of size
$N < N_M$ and the other half form within clusters $N > N_M$. For the
solar neighborhood surveys (Lada \& Lada 2003, Porras et al. 2003),
this median point has the value $N_M \approx 300$ (Adams et al. 2006).
When the distributed population is included, however, the median point
of the probability distribution moves downward to $N_M \approx 100$
(Allen et al. 2007).  

[2] It is generally accepted that only about 10 percent of the stellar
population is born within systems that are destined to become open
clusters, which are gravitationally bound over longer timescales of
100 -- 500 Myr (e.g., van den Bergh 1981, Elmegreen \& Clemmens 1985,
Battinelli \& Capuzzo-Dolcetta 1991, Adams \& Myers 2001). Suppose
that the probability distribution for star-forming units has equal
weight for each decade in $N$, as would be the case if the
distribution in the solar neighborhood extends upward as described
above. In this case, half of the stellar population would be born
within systems with $N \ge 1000$, but at most 20 percent of that
population would end up living in open clusters. In this case, the 
remaining 80 percent of the stars would be born within ``clusters''
that dissolve quickly, after only $\sim$10 Myr, a timescale that is
shorter than the relaxation time for these systems. As outlined below,
this timescale is also shorter than that required to explain many
Solar System properties. As a result, some evidence points toward the
Sun being born within a gravitationally bound cluster, as opposed to a
short-lived aggregate, and this requirement is realized for only about
10 percent of the stellar population.

In light of these uncertainties, the distribution $df_c/dN$ of cluster
membership sizes should be considered preliminary. More data are
required to clear up these uncertainties.

\subsection{Cluster Properties} 

In addition to the issue of what cluster size distribution $df_c/dN$
($dP/dN$) applies, several other issues arise. In particular, the
stellar membership size $N$ is not the only relevant variable.
Clusters with the same $N$ can have varying radii and hence varying
mean densities.  On a finer scale of distinction, clusters with the
same size $N$ and radius $R$ can have different distribution functions
for the stellar velocities, different stellar density profiles, or
different background potentials given by their gaseous component. In
the past several years, observations have started to place constraints
on the possible ranges of the cluster parameters, especially for the
clusters in the solar neighborhood (those within $\sim2$ kpc of the
Sun). A brief overview of these properties can be summarized as
follows (see Allen et al. 2007; Gutermuth et al. 2005, 2009; Lada \&
Lada 2003; Porras et al. 2003; and references therein):

The mean cluster radius $R$ scales with cluster membership size $N$ 
according to the power-law relation 
\be
R = R_0 \left( {N \over N_0} \right)^{\alpha} \, . 
\label{radlaw} 
\ee
For clusters in the solar neighborhood (Lada \& Lada 2003, Porras et
al. 2003), this law holds with parameters $R_0$ = 1 pc, $N_0$ = 300,
and $\alpha \approx 1/2$ (Adams et al. 2006). This relationship
defines the mean radius for a given $N$; the data show a scatter on
either side of this mean value with an amplitude of approximately a
factor of 2. For the full range of clusters, extending up to $N =
10^6$, equation (\ref{radlaw}) predicts overly large radii for
large-$N$ clusters if we use index $\alpha = 1/2$ (compare with data
presented in Chandar et al. 1999, Pfalzner 2009; see also Proszkow
2009).  Over this full range of cluster membership sizes, indices 
in the range $\alpha \approx 1/4 - 1/3$ provide a better fit.

With the cluster radii specified through equation (\ref{radlaw}), the
characteristic mean stellar density $\nchar$ is given by
\be
\nchar = {3 N \over 4 \pi R^3} \propto N^{1 - 3 \alpha} \, . 
\ee
For $\alpha$ = 1/2 (1/4), the stellar density decreases (increases)
with membership size $N$. In either case, however, the mean stellar
density is a relatively slowly varying function of $N$. Note that the 
intermediate value $\alpha$ = 1/3 leads to a constant stellar density.
Further, the typical mean value of the stellar density is of order 
$\nchar \sim 100$ pc$^{-3}$. This density affects the interaction
rates, and hence the probability that the early Solar System
suffered a close encounter with a passing star (see below).

Observations of cluster-forming molecular cloud cores show that the
gas density profiles have the approximate form $\rho \sim 1/r$ so that
the enclosed mass $M(r) \sim r^2$ (Larson 1985, Jijina et al. 1999).
Similarly, N-body simulations of young embedded clusters show that the
stellar number density also has a power-law form $n_\ast \sim 1/r^p$,
with index $p$ close to unity (e.g., Kroupa 1995).  For purposes of
this discussion, we need to estimate the probability that the Solar
System resides at a given radial location within a cluster. We can
thus use the probability distribution 
\be
{dP \over dr} = {4 \pi r^2 \over N} n_\ast(r) = {2 r \over R^2} \, , 
\label{dpdr} 
\ee
where $R$ is the cluster radius and where the distribution is
normalized so that $\int_0^R (dP/dr)dr$ = 1. The probability
distribution $dP/dr$ thus vanishes outside the cluster where 
$r > R$. The corresponding expectation value for the radial position
is given by $\langle r \rangle = 2R/3$, and the median radius is given
by $r_h = \sqrt{2} R/2$.  This form is valid for young clusters. At 
later times, the density profile is expected to become steeper and
approach the form $n_\ast \sim r^{-2}$, so that $dP/dr \approx 1/R$; 
for this case, $\langle r \rangle = R/2 = r_h$.  

\subsection{Cluster Dynamics} 

The relaxation time $t_R$ defines the clock that sets the pace for
cluster dynamics (Binney \& Tremaine 1987). For embedded clusters that
contain a substantial gaseous component, the relaxation time is given
approximately by
\be
t_R \approx {v \over R} {N \over 10 \epsilon^2 \ln(N/\epsilon) } \, , 
\label{trelax} 
\ee
where $v/R$ is the crossing time, $N$ is the cluster membership size,
and $\epsilon$ = $N \langle m_\ast \rangle/M_C$ is the star formation
efficiency; here, $\langle m_\ast \rangle$ is the mean stellar mass
and $M_C$ is the total cluster mass, including gas (Adams \& Myers
2001).  While the cluster retains its gaseous component, the
relaxation time is thus longer than that of purely stellar systems
(Binney \& Tremaine 1987). The behavior of the gas content thus plays an important
role in early cluster evolution (the first $\sim5$ Myr).

Massive stars sink to the cluster center over a relatively long time
scale, given approximately by $t_{R}/m$, where $t_{R}$ is the
dynamical relaxation time (e.g., equation [\ref{trelax}]) and $m$ is
the mass of the star in solar masses (Portegies Zwart 2009). Of course, 
the massive stars can also be formed at the cluster centers, and some
observational evidence (Testi et al. 2000) and theoretical considerations
(Bonnell \& Davies 1998) support this point of view. On the other hand, mass
segregation can be sped up through a combination of subvirial starting
velocities and sufficiently clumpy initial density distributions
(Allison et al. 2009, Moeckel \& Bonnell 2009). In any case, massive
stars are expected to reside near cluster centers.

For gravitationally bound systems, where about 10 percent of the
stellar population is born, the total lifetime $t_T$ of a cluster is
expected to be a multiple $Q$ of the relaxation time $t_R$ (Binney \&
Tremaine 1987), where $Q$ = 10 -- 100.  If we use the definition of
$t_R$, the virial relation $v^2 = GM/R$, and the number versus radius
relation of equation (\ref{radlaw}), the expected cluster lifetime
scales according to $t_T \sim N^{3/4}/\ln N$, where we have used
$\alpha$ = 1/2. For comparison, the timescale over which observed star
clusters are dissolved obeys an empirical law of the form $t_T$ = 2.3
Myr $M^{0.6}$, where the cluster mass $M$ is given in solar masses
(Lamers et al. 2005).  For $M$ = 300, this empirical relation implies that
$t_T \approx$ 70 Myr. Since the relaxation time for this type of
cluster is $t_R \approx$ 5 Myr, we get agreement for $Q \approx 14$.

In any case, clusters live for timescales that are much shorter than
the current age of the Solar System. It is interesting to determine
how many orbits the Solar System has made around the galactic center
since its birth. The circular velocity of the Sun around the Galaxy is
$v_{\rm cir} \approx 235$ km s$^{-1}$.  If we assume that the orbit
speed and galactocentric radius $R_{g}$ have not changed, this
exercise implies that the solar system has made about $N_{orb}$ =
$(v_{\rm cir} t)/(2 \pi R_{g}) \approx$ 22 orbits.  The Solar System
has thus traveled an enormous distance (more than one Mpc) since its
birth.  Note that the Solar System is likely to have experienced many
(wide) encounters that change its velocity vector during the course of
its lifetime. As outlined in the following section, however, the Solar
System is unlikely to have experienced close encounters with any
passing stars, as such perturbations would have left dramatic --- and
unobserved --- signatures in our planetary orbits.

\section{CONSTRAINTS FROM DYNAMICS} 

\subsection{Encounter Rates in Embedded Clusters} 

Within a cluster, the rate $\Gamma$ at which a given solar system
encounters other stellar members can be written in the form 
\be
\Gamma = \langle n \sigma v \rangle \, , 
\ee
where $n$ is the number density of potential target systems, $\sigma$
is the cross section for the given interaction, and $v$ is the typical
speed at which the solar system orbits through the cluster. The
velocity and number density vary with time and with position in the
cluster, so that averaging is necessary, as indicated by the angular 
brackets. In addition, interaction cross sections depend on the 
encounter speeds.

For a given cluster, we define $\Gamma(b)$ to be the rate at which a
given solar system encounters passing stars within a distance $b$.
Numerical (N-body) simulations show that this encounter rate can be
written in the form 
\be
\Gamma = \Gamma_0 \left( {b \over b_0} \right)^\exgamma \, , 
\label{rate} 
\ee
where $\Gamma_0$ and $\exgamma$ are constants, $b$ is the distance of
closest approach, and $b_0$ is a reference distance scale (Proszkow
\& Adams 2009). In the absence of gravitational focusing, the index 
$\exgamma \approx 2$; in practice, one finds somewhat smaller values
in the range $\exgamma = 1 - 2$, where $\exgamma$ decreases slowly
with the range of close encounters under consideration.  Without loss
of generality, we can take $b_0$ = 1000 AU. The constant $\Gamma_0$
depends on the specifics of the cluster properties; for the parameters
expected for possible solar birth clusters, $\Gamma_0$ typically lies
in the range $\Gamma_0 = 0.01 - 0.1$ encounters per star per Myr.
These values can be understood as follows: For clusters in the
present-day solar neighborhood, the mean stellar density is about 
$n_0 \approx 100$ pc$^{-3}$, and the typical velocity dispersion 
$v_0 \approx 1$ km s$^{-1}$.  The nominal value of $\Gamma_0 \sim n_0
v_0 b_0^2$ is thus of order $\Gamma_0 \sim 0.0025$ Myr$^{-1}$. The
actual value of $\Gamma_0$ is larger because the clusters start with
subvirial conditions and spend much of their embedded phase with
larger densities; the clusters thus contract by a factor of about
$\sqrt{2}$ in radial scale and hence a factor of about $2\sqrt{2}$ in
density. In addition, the interactions are more frequent in the
cluster core, where the density is higher, and this effect also
increases $\Gamma_0$.

For a given timescale $t_C$, the above considerations define a
characteristic distance of closest approach, denoted here as $b_C$.
By setting $\Gamma t_C$ = 1 in the interaction rate of equation
(\ref{rate}), we find the characteristic distance scale $b_C$ for
close encounters over that span of time, i.e., 
\be
b_C = b_0 (\Gamma_0 t_C)^{-1/\exgamma} \, . 
\label{typicald}
\ee 
Note that the lifetime of embedded clusters (Allen et al. 2007), the
expected lifetime of circumstellar disks (Hern{\'a}ndez et al. 2007),
and the time required to form giant planets (Lissauer \& Stevenson
2007) are of order 3 -- 10 Myr. For the sake of definiteness, we set
$t_C$ = 10 Myr and find that $b_C \approx 300 - 1000$ AU. In this 
sense, the ``typically expected'' distance of closest approach
experienced by the early solar system is several hundred AU (Bonnell
et al. 2001, Adams et al. 2006, Malmberg et al. 2007).  For other
parameter choices, the distance scale $b_C$ can be found using
equation (\ref{typicald}).

\subsection{Orbital Considerations} 

Another relevant property of clusters is their distribution of orbits.
Compare the extreme cases of purely radial orbits and purely circular
orbits: For radial orbits, solar systems pass through (or near) the
cluster center every crossing time ($\sim1$ Myr). The cluster center
is the densest region, and contains the most massive stars, so radial
orbits lead to maximal disruption in terms of both radiation exposure
and probability of scattering encounters. In contrast, for a given
orbital energy, circular orbits allow solar systems to stay as far as
possible from the cluster center and thereby minimize the probability 
of disruption.  
 
One standard way to characterize the orbits in a dynamical system is
to define the parameter $\beta$ according to 
\be
\beta \equiv 1 - { \langle v_\theta^2 \rangle \over 
\langle v_r^2 \rangle } \, , 
\ee 
where $v_r$ and $v_\theta$ are the radial and poloidal components of
the velocity (Binney \& Tremaine 1987). Isotropic distributions of
velocity lead to $\beta$ = 0, whereas radial orbits result in $\beta$
= 1.

Observations of young embedded clusters are starting to provide clues
to the expected values of $\beta$. Growing observational evidence
suggests that forming stars in clusters are not born with virial
velocities (Walsh et al. 2004, Peretto et al. 2006).  Instead, the
clumps that collapse to form stars move more slowly through the
system, and only begin to move ballistically after star/disk formation
is complete.  With these subvirial starting conditions, the initial
orbits are directed more radially inward (compared with virial
starting states) and some dynamical memory of this initial condition
is retained (Adams et al. 2006). The loss of the gaseous component in
young clusters can induce an additional radial component to the
stellar velocities.  As a result, subvirial starting conditions lead
to $\beta \approx 1/2$, a value intermediate between radial and
circular orbits. On a related front, kinematic observations of stars
in young clusters are now possible, and recent data indicate that the
Orion Trapezium Cluster displays the kinematic signatures of subvirial
starting conditions (Tobin et al. 2009, Proszkow et al. 2009).  These
observational considerations, while indirect, suggest that forming
clusters generally have moderately radial velocity distributions.
This finding, in turn, increases the probability of solar system
disruption through both radiation exposure and scattering
encounters. For the future, it would be useful to have further
observational specification of cluster velocity distributions to help
assess environmental effects.

\subsection{Disk Truncation} 

During the early stages of solar system formation, close encounters by
passing stars can disrupt the solar nebula and thereby limit the mass
reservoir available for planet formation (Clarke \& Pringle 1993,
Ostriker 1994, Heller 1995, Kobayashi \& Ida 2001; see also Kenyon \&
Bromley 2001 for a discussion of the effects of gravitational
stirring). Taken together, these studies show that passing stars act
to truncate circumstellar disks during close encounters. If $b$ is the
impact parameter of the encounter, the disks are generally truncated
at a radius $r \approx b/3$. Most of the material outside this
truncation radius is either left unbound, or is captured by the
passing star. In addition, the surface density of the remaining disk
(inside the truncation radius $r = b/3$) is perturbed during the
encounter.

As outlined in Section 2.5, the inferred gas surface density of the
early solar nebula has a relatively sharp edge at $r \approx 30$ AU.
This estimate is based on reconstituting the nebula based on the
masses, compositions, and orbits of the planets. Although this
procedure is not without uncertainties, the edge at approximately 30
AU remains a robust result. This finding constrains any close
encounters that took place while the early solar nebula was intact to
have impact parameters $b \ge 90$ AU. Since the timescale for giant
planet formation (Lissauer \& Stevenson 2007) and the expected
lifetimes of circumstellar disks (Hern{\'a}ndez et al. 2007) are both
about 10 Myr, this constraint applies over the first $\sim10$ Myr of
Solar System history.

Using equation (\ref{rate}), we can estimate the rate at which the
early Solar System experienced close encounters with $b < 90$ AU: 
This rate is expected to fall in the range $\Gamma_{90} \sim 10^{-4} -
10^{-2}$ encounters per Myr. For purposes of this discussion, we take
the encounter rate to be $\Gamma_{90} = 10^{-3}$ Myr$^{-1}$.  Over the
fiducial timescale of 10 Myr required for planet formation (and
observed disk lifetimes), the probability of such a close encounter is
low, with odds of only about 1 in 100. Even in more interactive
clusters, the odds are only 1 in 10. Although the interaction rates
can vary with cluster properties, these results suggest that the early
solar nebula in unlikely to have been truncated so severely that giant
planet formation is compromised.

Recent observations have begun to constrain the form of the surface
density for typical circumstellar disks. For example, one study finds
that disks in the Rho Ophiuchus star forming region have surface
densities of the approximate form $\Sigma \sim r^{-p} \exp[-r/r_0]$,
where the scale $r_0$ = 20 -- 200 AU (Andrews et al. 2009; see also
Isella et al. 2009). Although preliminary, these results suggest that
unperturbed disks might have edges that are softer than the edge
structure inferred for the solar nebula (Section 2.5).  For example,
in our Solar System the observed Kuiper Belt has a mass of $\sim 0.08$ 
$M_E$ (Luu \& Jewitt 2002); if we augment this mass by a factor
of 10 for mass loss, and another factor of 100 to add back in the
gasecous component, the inferred surface density is smaller than that
of the MMSN (at 30 AU) by a factor of $\sim 40$. This density contrast 
occurs abruptly, whereas an exponential fall-off requires 3.7 scale
radii $r_0$ (74 -- 740 AU) to produce such a large decrease in surface
density. As a result, some type of truncation event may be required to
explain the observed (apparent) edge of the Solar System. In addition
to close encounters, the nebula can be truncated by photoevaporation,
as discussed in the following section. Note that both mechanisms arise
from the background environment.

For completeness, we note that the cluster environment can also add
mass to the early solar nebula. As the star/disk system orbits through
the cluster potential, the nebula can gain mass through Bondi-Hoyle
accretion (Throop \& Bally 2008). Through this process, the feeding zone of the
solar nebula is given by $R_{BH} = 2 G M_\ast /(v^2 + c_s^2)$, where
$v$ is the system speed with respect to the cluster and $c_s$ is the
sound speed of the background gas. For typical cluster parameters, the
rate of mass accretion ${\dot M} = \pi R_{BH}^2 \rho v$ falls in the
range ${\dot M}$ = $10^{-8} - 10^{-9} M_\odot$ yr$^{-1}$. Since gas is
typically retained in these clusters for only 3 -- 5 Myr (Gutermuth et
al. 2009, Allen et al. 2007), the amount of mass added to the disk is
expected to fall in the range $(\Delta M)_d \approx 0.003-0.05 M_\odot$. 
Under favorable circumstances, the cluster environment can thus
provide the early solar nebula with a mass comparable to the MMSN.

\subsection{Disruption of Planetary Orbits}

After the planets have formed, and disk truncation is no longer an
issue, the planetary orbits are susceptible to disruption.  As
outlined in Section 2, the orbits of the giant planets in our Solar
System are remarkably well-ordered, with low eccentricities and a
narrow spread in inclination angle. Although the Solar System would
not be seriously compromised if the eccentricities or inclination
angles were somewhat higher, we can use these properties to constrain
possible interactions between the early Solar System and other stars
in its birth cluster. If we let $\cross$ denote the cross section for
disrupting the orbits of the giant planets, this constraint can be
written in the general form
\be 
\int n_\ast \cross v dt < 1 \, , 
\label{scattercon} 
\ee 
where the $n_\ast$ is the density of passing stars, $v$ is the
relative speed, and the integral is taken over the time spent in the
birth cluster. Note that additional interactions could, in principle,
take place after the Solar System leaves its birth cluster. However,
the density of passing stars is lower and the interaction cross
sections are smaller (because of the higher relative velocities), and
these trends more than compensate for the longer available time.

To evaluate the scattering constraint of equation (\ref{scattercon}),
we consider the Solar System to be ``disrupted'' if the eccentricities
of the giant planet orbits are doubled, or if the spread in their
inclination angles is doubled.  Notice that this level of disruption
is not severe, in that the Solar System could have functioned with
larger orbital eccentricities or inclination angles. However, the 
well-ordered nature of the current planetary orbits shows that such 
disruption did not in fact take place and we can use this property 
to place limits on the dynamical history of the Solar System 
(in the absence of a strong damping mechanism for orbital 
eccentricity and inclination). 

The cross sections for planetary disruption has been calculated
through an extensive series of Monte Carlo simulations (Adams \&
Laughlin 2001; see also Heggie \& Rasio 1996, Bonnell et al. 2001,
Adams et al. 2006, Malmberg \& Davies 2009, Spurzem et al. 2009).  In
these calculations, we consider the reduced Solar System consisting of
the four giant planets and the Sun, where the planets have their
measured masses and semi-major axes, but zero eccentricity, and all
orbits lie in the same plane. These solar systems are then subjected
to fly-by encounters with passing binary stars, where the binary
properties, the parameters of the encounter (e.g., impact parameter),
and the phases of the planetary orbits are sampled according to a
Monte Carlo scheme.  The results of these simulations are then used 
to construct the cross sections for varying levels of solar system 
disruption.

\begin{figure} 
\centerline{\epsscale{0.90} \plotone{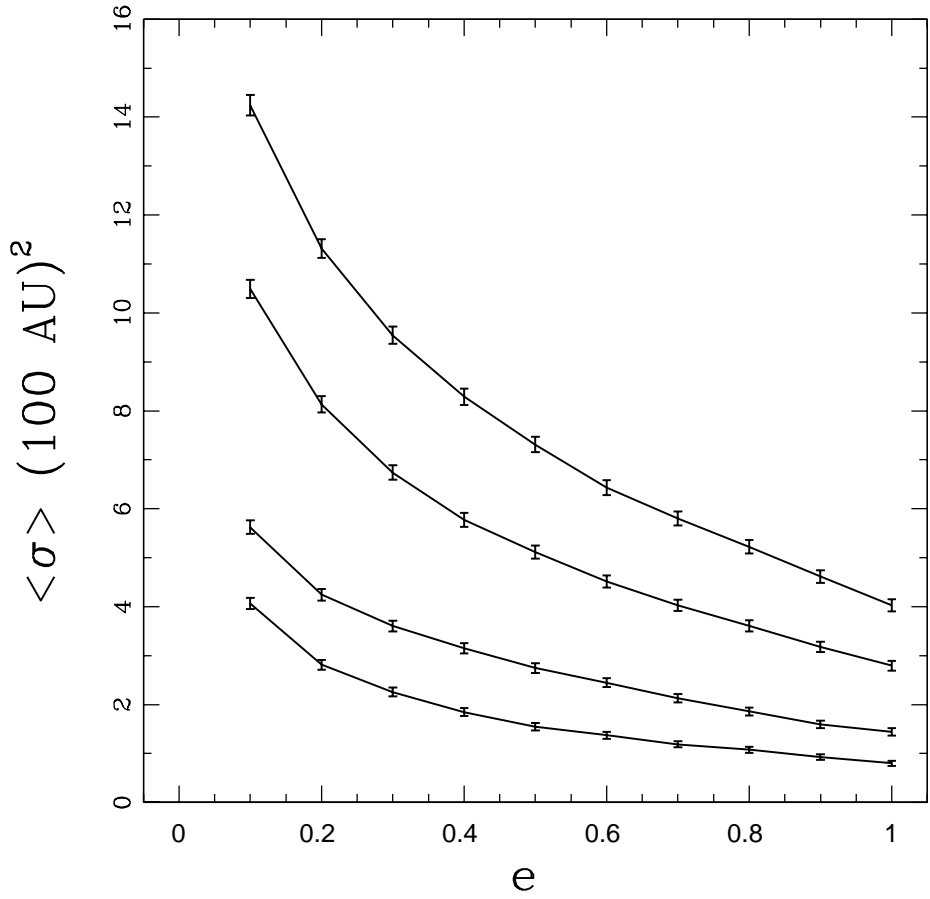}}
\caption{Cross sections for disrupting the Solar System through close 
encounters with passing binary stars. The cross sections are given
here for increasing the eccentricity of each planet to a given value
$e$, presented here as a function of eccentricity $e$. The four curves
correspond to the four giant planets from Neptune (top) to Jupiter
(bottom). The error bars show the uncertainty due to incomplete
sampling in the Monte Carlo procedure (adapted from Adams \& Laughlin
2001). }
\label{fig:cross} 
\end{figure}

Figure \ref{fig:cross} shows the cross sections for impulsively
increasing the eccentricities of the four giant planets, given here 
as a function of eccentricity. The cross section for increasing the 
eccentricity beyond unity (right side of the figure) corresponds to
ejection of the given planet. The cross section for disrupting the
solar system (according to the criteria outlined above) takes the
value 
\be 
\cross \approx 160,000 ({\rm AU})^2 \, \approx 
4 \times 10^{-6} ({\rm pc})^2 \, .
\label{csection} 
\ee
This cross section corresponds to a distance of closest
approach $b \sim 225$ AU.

In recent models of the early Solar System, the giant planets can form
in a more compact configuration and then slowly migrate outwards
(Tsiganis et al. 2005). Although the tighter orbits could result in a
somewhat smaller initial cross section, the difference is small (only
a factor of $\sim2/3$ since the cross sections tend to scale linearly
with the outermost semimajor axis; Adams \& Laughlin 2001). In
addition, eccentricity increases in such a compact configuration can
lead to more disruption in subsequent evolution (Malmberg et al.
2007). As a result, equation (\ref{csection}) provides a good working
estimate for the disruption cross section.

To interpret this result, we first note that the dynamical speed $v$
within these clusters is typically $v \sim 1$ km s$^{-1}$ $\approx$ 
1 pc Myr$^{-1}$.  If we write the number density of stars $n_\ast$ in 
units of pc$^{-3}$ and the time $t$ in Myr, the constraint from the
required survival of planetary orbits takes the form 
\be
\int n_\ast \, dt \, < 250,000 \, {\rm Myr} \, {\rm pc}^{-3} \, . 
\label{survivenep}
\ee
This constraint shows that even for a relatively high density in the 
star cluster, say $n_\ast \sim 1000$ pc$^{-3}$, the Solar System can
survive for $\sim$250 Myr before disruptions become likely. A similar
constraint follows from requiring that the inclination angles of the
orbits of Neptune and Uranus are not overly perturbed by passing stars
(Gaidos 1995). Note that these (long) timescales can only be realized  
if the Sun forms within a long-lived cluster, which occurs for about 
10 percent of the stellar population. 

\subsection{Smaller Solar System Bodies}

The discovery of an extended scattered disk in the Solar System
(Gladman et al. 2002), including the trans-Neptunian object Sedna
(Brown et al. 2004), could represent additional evidence for
scattering interactions between our Solar System and passing
stars. The orbit of Sedna has a large eccentricity ($e \approx 0.84$)
with perihelion $p = a (1 - e)$ $\approx$ = 70 AU. This orbit is thus
rather unusual among solar system bodies. Recent numerical studies
have shown that this orbit can be produced by a close encounter with a
passing star. In one model, Sedna initially resides in the scattered
disk and has its perihelion lifted by the encounter (Morbidelli \&
Levison 2004; Brasser et al. 2006); in another model, the encounter
scatters Sedna from the Kuiper belt into its observed eccentric orbit
(Kenyon \& Bromley 2004).  The required impact parameter for such a
collision lies in the range $b \approx$ 400 -- 800 AU.  A close
encounter at approximately this distance could also help account for
the observed edge of the Kuiper Belt at $r \sim 50$ AU.  However, it
is difficult to produce a clean edge with a distant fly-by encounter,
so that explaining the observed edge would indicate a somewhat closer
encounter with approach distance $b$ = 200 -- 300 AU (see Levison et
al. 2004, Kenyon \& Bromley 2004, Adams \& Laughlin 2001).  As
outlined in Section 4.1, a typical solar system in a typical cluster
is expected to experience about one such close encounter over a time
span of 10 Myr. The finding that our Solar System requires (of order)
one such encounter is thus consistent with the idea that the Sun
formed within a moderate-sized cluster. More specifically, requiring
at least one encounter with $b \le 400$ AU implies that $\int n_\ast
dt \ge$ 80,000 Myr pc$^{-3}$ (compare with equation
[\ref{survivenep}]). Meeting this constraint is more likely if the Sun
forms within a gravitationally bound cluster, which occurs about 10
percent of the time.

We also note that any stellar encounter that led to the creation of an
effective edge to the Kuiper Belt must occur sufficiently early in the
Solar System's evolution. If the encounter takes place more than 10
Myr after the Oort cloud begins forming, either the scattered disk
contains too many bodies or the Oort cloud is compromised (Levison et al. 2004). 

Before leaving this section, we note that solar systems can eject a
large number of rocky bodies during their early phases of evolution.
Although this rocky material represents a small fraction of the total
mass in solids, about 10 percent, the number of rocks is expected to
be large, perhaps $N_R \sim 10^{16}$ bodies with mass $m > 10$ kg.
Some fraction of this ejecta remains bound to the cluster and can be
captured by other solar systems residing in the birth aggregate (Adams
\& Spergel 2005, Belbruno et al. 2008).  The cross section for rock
capture by binary systems is also large, at least $\langle \sigma
\rangle \sim (200 \, {\rm AU})^2$ for the typical velocity dispersions
of clusters. With these parameters, every solar system in a cluster
can share rocks with every other solar system in the same cluster. One
implication of the Sun forming in a cluster is that the Solar System
is likely to contain rocks that originated in many other solar systems
(perhaps thousands).

\section{CONSTRAINTS FROM RADIATION FIELDS}

\subsection{Radiative Processes in Embedded Clusters}  

In cluster environments, the ultraviolet (UV) radiation provided by
the background often dominates that provided by the central star. Such
energetic radiation leads to photoevaporation of circumstellar disks
and hence loss of planet forming potential.  If our Solar System
formed within such a cluster, the outer edges of the early solar
nebula could be truncated by evaporation. As outlined above, the
nebula extended out to at least $r \sim 30$ AU, near the current
semi-major axis of Neptune's orbit. The solar birth cluster is thus
constrained --- the background radiation must be weak enough to allow
gas to survive in the solar nebula at radii $r \le 30$ AU.

Two types of radiation are important in this context: Ionizing photons
with $E_\gamma = h \nu \ge 13.6$ eV, known as EUV radiation, are most
efficient at driving photoevaporation (on a per photon basis).  The
other radiation band of interest, known as FUV, corresponds to the
next lower range of energy, where 6 eV $\le h \nu <$ 13.6 eV.  In the
largest clusters, EUV radiation is generally the most important.
However, EUV radiation is primarily emitted by the largest stars,
spectral type OB, which are rare. As a result, in moderate-sized
clusters, the photoevaporation process due to external radiation is
often dominated by FUV radiation (Hollenbach et al. 1994, Johnstone et
al. 1998, St{\"o}rzer \& Hollenbach 1999, Armitage 2000).

\subsection{Distribution of UV Radiation Fields} 

Clusters provide a wide range of possible UV fields that can affect
forming solar systems. The most extreme environments are too hostile
for planet formation to take place. As a result, our own planetary
system is constrained to have formed in the presence of more moderate
radiation fields.

Specification of the radiation fields in clusters involves three
distributions: First, clusters come in different membership sizes $N$,
so the distribution of clusters $df_c/dN$ comes into play (see Section
3.1). Second, for a fixed value of $N$, different realizations of the
stellar IMF ($N$ times per cluster) lead to a distribution of total UV
luminosities for clusters with a fixed membership size $N$. Finally,
photoevaporation depends on UV flux, rather than UV luminosity, and
the flux depends on the position of the target solar system within the
cluster. Photoevaporation thus depends on the distribution of radial
positions of the stars within their cluster, and this distribution is
related to the mass profile of the cluster.

If the stellar IMF is fully sampled, the UV luminosity can be
characterized by its mean value (where the meaning of being fully
sampled is clarified below).  Here UV refers to either the EUV or the
FUV band. We thus define 
\be
\langle L_{UV} \rangle_\ast \equiv \int_0^\infty L_{UV}(m) 
{d N_\ast \over dm} dm \, , 
\ee 
where the stellar IMF $dN_\ast/dm$ vanishes for masses above the upper
cutoff $m > m_\infty$ and below the brown dwarf limit $m < m_{min}$.
This expectation value is normalized so that $\langle L_{UV}
\rangle_\ast$ corresponds to the mean UV luminosity per star. This
quantity is defined once the stellar IMF is specified (assuming that
stellar structure models adequately determine the UV fluxes for a
given mass). Since the UV luminosity is dominated by the high mass
stars, and these objects evolve to the main sequence quickly, we can
use stellar configurations on the zero-age main sequence to evaluate
this expectation value.  For the usual IMF (slope $\gamma$ = 2.35,
$m_\infty$ = 100 -- 120), these mean values are $\langle L_{FUV}
\rangle_\ast \approx 1.3 \times 10^{36}$ erg s$^{-1}$ and $\langle
L_{EUV} \rangle_\ast \approx 9.3 \times 10^{35}$ erg s$^{-1}$
(Armitage 2000, Fatuzzo \& Adams 2008).

To start, we ignore the fact that the UV luminosity in a cluster will
have a fairly wide distribution.  The expectation value of the UV
luminosity for a cluster of membership size $N$ is given by 
\be
L_{UV} (N) = N \langle L_{UV} \rangle_\ast \, , 
\label{luvmean} 
\ee
where the subscript UV refers to either the FUV or EUV wavelength
band. In the limit of large $N$, this expectation value provides a
good estimate, and the distribution of luminosity about this mean
value approaches a gaussian form due to the central limit theorem. 
In practice, however, convergence is extremely slow and the
moderate-sized clusters of interest display large departures from
gaussianity. In particular, the median values of the distributions 
are significantly below the mean (Fatuzzo \& Adams 2008) and the 
distributions are wide. For example, the standard deviation of the
distribution of $L_{FUV}$ is larger than the mean value (given by
equation [\ref{luvmean}]) for $N \le 700$. Similarly, the standard
deviation of the $L_{EUV}$ distribution is wider than the mean for 
$N \le 1400$. These results indicate that the cluster to cluster 
variation (for fixed $N$) is important for moderate-sized systems.
More specifically, for the solar neighborhood cluster sample (Lada 
\& Lada 2003, Porras et al. 2003), half of the stars are found in 
clusters with $N \le 300$; and $\sim80$\% (90\%) of the stars are
found in clusters with $N \le 700 (1400)$.

\begin{figure} 
\centerline{\epsscale{0.90} \plotone{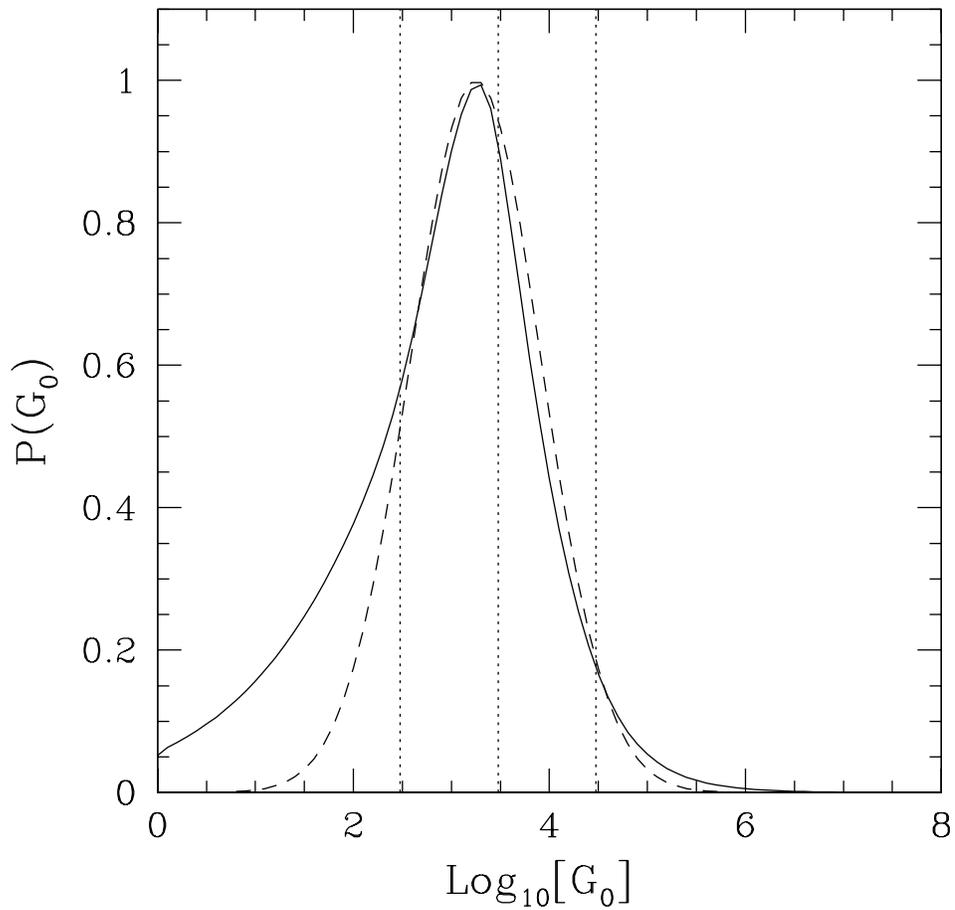}}
\caption{Distribution of FUV radiation fluxes for the collection of young
embedded clusters in the solar neighborhood. This distribution is
obtained by convolving the distribution of cluster membership sizes
$N$, the distribution of FUV luminosities for fixed $N$ due to different 
sampling of the stellar initial mass function, and the distribution of 
radial positions with the clusters. The three vertical lines delimit 
benchmark values of $G_0$ = 300, 3000, and 30000 (adapted from 
Adams et al. 2006). } 
\label{fig:fuvdist} 
\end{figure}

The above considerations describe the expected values of UV
luminosities for cluster environments. To understand the possible
impact on the early Solar System, however, we must determine the
distribution of UV fluxes. We focus here on the case of the FUV flux
distribution; the EUV fluxes can be treated in similar fashion.  This
overall distribution depends on three input distributions: First, the
clusters themselves come in a variety of membership sizes $N$. Second,
for a given $N$, each cluster will sample the stellar IMF differently,
and the resulting sampling will give rise to a distribution of UV
luminosity for fixed $N$ (see above).  Finally, for a given cluster
size $N$ and realization of the IMF, the distribution of radial
positions within the cluster produces a corresponding distribution of
UV fluxes. The resulting composite FUV flux distribution, for the
collection of clusters found in the solar neighborhood, is shown in
Figure \ref{fig:fuvdist}.  The FUV fluxes are expressed in units
$G_0$, defined so that $G_0$ = 1 corresponds to the value $F_{FUV}$ =
$1.6 \times 10^{-3}$ erg s$^{-1}$ cm$^{-2}$ (close to the value
appropriate for the interstellar medium; see Habing 1968). The cluster
environment thus produces FUV radiation fluxes that are thousands of
times more intense than in the field. Similar results hold for the
case of EUV radiation (Armitage 2000).

We can gain further insight by considering the expected flux levels
within clusters with given membership $N$ (as a function of $N$). The
mean luminosities for both FUV and EUV radiation are given above. For
clusters with exceptionally large $N$, the expected luminosity is
close to mean; for more moderate clusters, however, the median
luminosity is smaller than the mean by a factor $f_m \approx 0.8$
(Fatuzzo \& Adams 2008). To estimate the flux, we must specify the
radial location, which we take to be the expectation value $r$ =
$\langle r \rangle = 2R/3$, where $R$ is the cluster radius (see
equation [\ref{radlaw}]).  The expected UV flux is thus given by
\be
F_{UV} \approx {f_m N \langle L_{UV} \rangle_\ast \over 4 \pi r^2} = 
{9 f_m N_0 \langle L_{UV} \rangle_\ast \over 16 \pi R_0^2} 
\left( {N \over N_0} \right)^{1 - 2 \alpha} \, . 
\ee
The fiducial flux level $F_{UV} = 9 f_m N_0 \langle L_{UV}
\rangle_\ast / 16 \pi R_0^2$ is thus $F_{FUV} \approx 5.8$ erg
s$^{-1}$ cm$^{-2}$ ($G_0$ $\approx$ 3600) for FUV, and $F_{EUV}
\approx 4.2$ erg s$^{-1}$ cm$^{-2}$ ($2 \times 10^{11}$ photons
s$^{-1}$ cm$^{-2}$) for EUV radiation. 

As outlined below, these flux levels (especially for FUV) are intense
enough to affect the early solar nebula.  Since the index $\alpha \sim
1/2$, as least for clusters in the solar neighborhood, these fiducial
flux levels are slowly varying with cluster membership size. On the
other hand, in the regime of large $N$, the cluster radius index
$\alpha = 1/4 - 1/3$, so that the flux levels are higher. In addition,
the inner portions of large clusters produce much stronger radiative
fluxes. However, the solar nebula can survive in such large clusters,
provided that the Sun spends much of its time in the outer regions. 
This point is reinforced in subsequent sections.

\subsection{Photoevaporation of Disks} 

When a disk is exposed to external UV radiation, the gas can be heated
to sufficiently high temperatures to drive an evaporative flow. This
process defines a critical fiducial length scale, the radius at which
the sound speed of the heated gas exceeds the escape speed from the Sun:
\be
r_g = {G M_\ast \over a_S^2 } = {G M_\ast \langle \mu \rangle \over kT} 
\approx 100 \, {\rm AU} \, \left( {T \over 1000 \, {\rm K}} \right)^{-1} \, , 
\ee 
where we have used the mass of the Sun. If EUV photons can penetrate
the outward flow with sufficient flux, they can heat the gas to
temperatures $T \sim 10^4$ K. On the other hand, FUV photons generally
heat the gas to lower temperatures with $T = 100 - 3000$ K. The
delineation of the regimes for which EUV and FUV radiation dominates
the mass loss is complicated (Johnstone et al. 1998, St{\"o}rzer \&
Hollenbach 1999, Armitage 2000, Adams et al. 2004, Clarke 2007,
Ercolano et al. 2009).  A brief overview is presented below.

For EUV radiation, the mass loss rate from a disk due to 
photoevaporation can be written in the approximate form 
\be
{\dot M} \approx (9 \times 10^{-8} \, M_\odot \, {\rm yr}^{-1}) 
\left( {\Phi \over 10^{49} \, {\rm s}^{-1} } \right)^{1/2} 
\left( {d \over 10^{17} \, {\rm cm}} \right)^{-1} 
\left( {r_d \over 30 \, {\rm AU} } \right)^{3/2} \, , 
\label{mdoteuv} 
\ee
where $\Phi$ is the EUV photon luminosity, $d$ is the distance of the
solar system to the cluster center, and $r_d$ is the disk radius
(Shu et al. 1993, Johnstone et al. 1998). The mass outflow rate thus 
scales according to ${\dot M} \propto F_{EUV}^{1/2}$, where $F_{EUV}$ 
is the flux of EUV photons from the cluster (keep in mind that the 
EUV is assumed to be generated by massive stars at the cluster center).

Consider a typical cluster in the solar neighborhood, with $N$ = 300
and radius $R$ = 1 pc. For a standard stellar IMF, the system is
expected to have 1 or 2 stars with mass $M_\ast > 10 M_\odot$.  If we
consider a typical solar system to lie at a distance $d$ = $R/2$, the
EUV flux is $F_{EUV} = \Phi/(4 \pi d^2)$ $\approx$ 4.3 $\times
10^{11}$ photons s$^{-1}$ cm$^{-2}$, the mass outflow rate from
equation (\ref{mdoteuv}) becomes ${\dot M} \approx 6.8 \times 10^{-9}
M_\odot$ yr$^{-1}$. This evaporation rate can be converted into a
timescale by assuming a starting disk mass, which we take to be 
$M_d$ = 0.05 $M_\odot$ (comparable to the MMSN). The resulting
fiducial timescale for evaporation (with disk radius $r_d$ = 30 AU) 
is $t \approx 147$ Myr, longer than typical disk lifetimes.  A solar 
system at a typical location in moderate-sized cluster is not greatly
affected by photoevaporation from EUV radiation. Larger EUV fluxes can
disrupt the disk.

\begin{figure} 
\centerline{\epsscale{0.90} \plotone{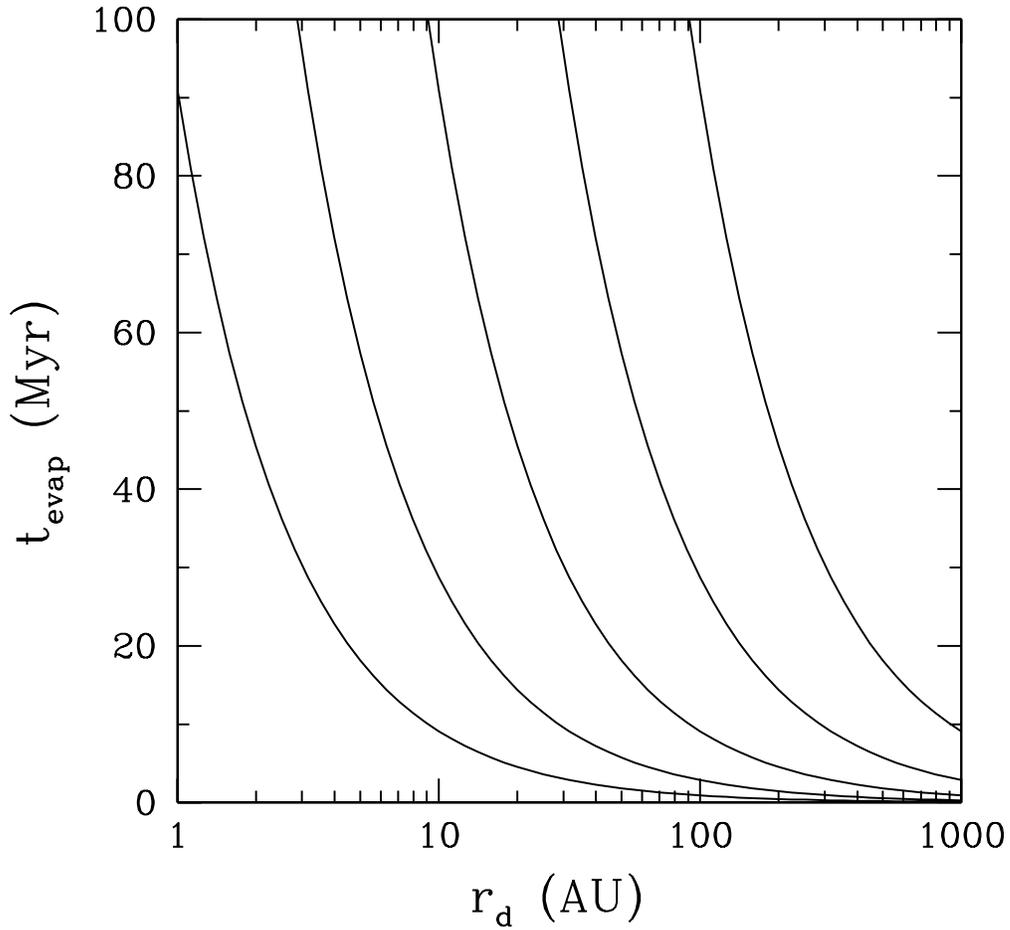}}
\caption{Photoevaporation timescales for the solar nebula as 
a function of disk radius $r_d$ for varying external EUV fluxes.
The five curves correspond to EUV fluxes $F_{EUV}$ = $10^{11}$, 
$10^{12}$, $10^{13}$, $10^{14}$, and $10^{15}$ photons s$^{-1}$ 
cm$^{-2}$ (increasing from right to left). Typically expected 
flux levels correspond to the lower values (see text). } 
\label{fig:euvtime} 
\end{figure}

To illustrate these trends, Figure \ref{fig:euvtime} plots the
timescale for disk evaporation as a function of outer disk radius. 
We assume that the disk mass scales with disk radius according to 
$M_d \propto r_d^{1/2}$, as expected for surface density $\Sigma
\propto r^{-3/2}$.  The five curves correspond to varying EUV fluxes
from $F_{EUV}$ = $10^{11}$ to $10^{15}$ photons s$^{-1}$ cm$^{-2}$. 
The lowest flux corresponds to that expected for a solar system living
at the edge of a moderate-sized cluster with $N$ = 300. The
evaporation timescales become problematic (where the solar nebula is
evaporated in 10 Myr at radius 30 AU) only when the EUV flux is
$\sim1000$ times the nominal value. Note that for less steep surface
density profiles (e.g., $\Sigma \propto r^{-1}$), the evaporation time
scales are even longer for $r_d < 30$ AU.

Given that EUV radiation becomes important only in extreme regimes of
parameter space, we now turn to the effects of FUV radiation. As shown
in Figure \ref{fig:fuvdist}, clusters in the solar neighborhood
provide a well-defined distribution of FUV fluxes, with typical values
in the range $G_0$ = 1000 to 10,000. For comparison, if we use the
expectation value for the FUV luminosity of a cluster with $N$ = 300
(see equation [\ref{luvmean}]), the expected flux at $R$ = 1 pc 
corresponds to $G_0 \approx$ 2000.  

\begin{figure} 
\centerline{\epsscale{0.90} \plotone{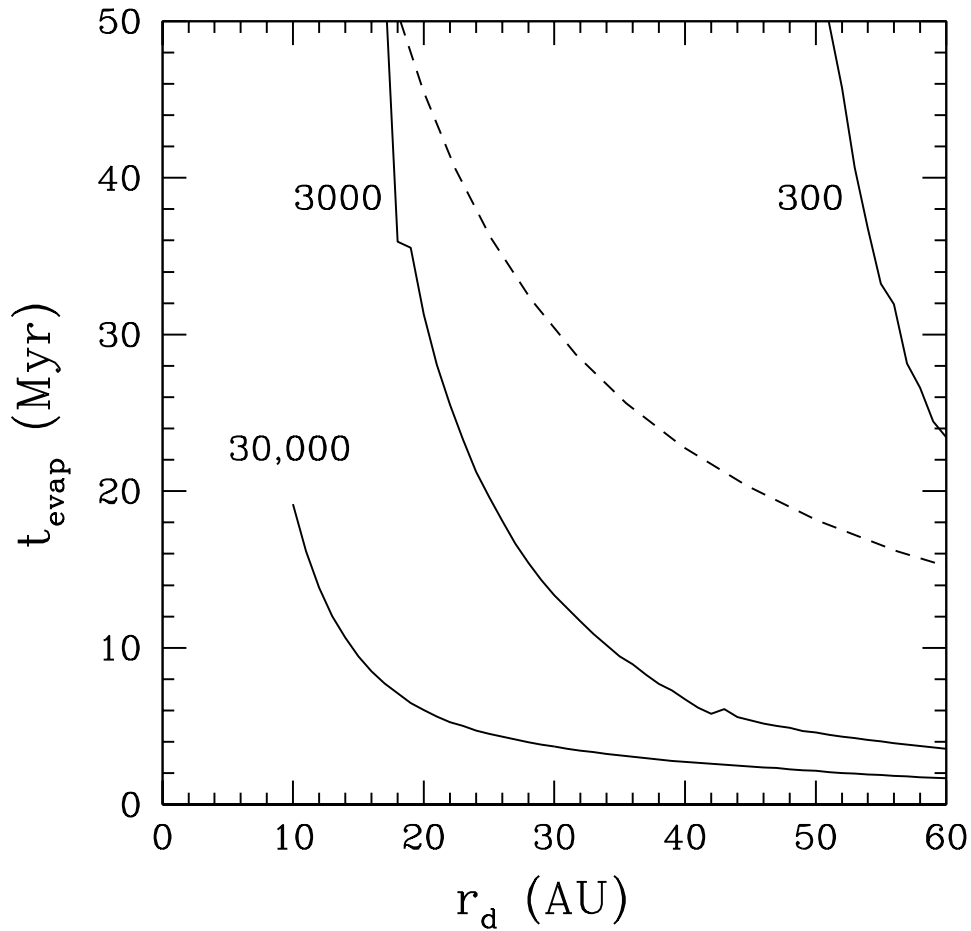}}
\caption{Photoevaporation timescales for the solar nebula as a 
function of disk radius $r_d$ for varying external FUV fluxes.  
The three solid curves correspond to FUV fluxes with $G$ = 300, 3000,
and 30,000 as labeled. The dashed curve shows the timescale for EUV
evaporation with EUV flux $F_{EUV}$ = $10^{13}$ photons s$^{-1}$
cm$^{-2}$, which corresponds to 100 times the ``typical value'' 
(see text).  } 
\label{fig:fuvtime} 
\end{figure}

Figure \ref{fig:fuvtime} shows the expected evaporation times for a
solar nebula heated by external FUV radiation (the mass loss rates are
taken from Adams et al. 2004, which uses the heating/cooling formalism
from Kaufman et al. 1999). The three solid curves correspond to the
benchmark values of FUV flux with $G_0$ = 300, 3000, and 30,000. Note
that this figure is plotted on a smaller scale than that used to
illustrate the effects of EUV radiation (compare with Figure
\ref{fig:euvtime}) because the expected FUV radiation can evaporate
the solar nebular down to smaller radii. For comparison, Figure
\ref{fig:fuvtime} also shows the timescales for evaporation with an
EUV flux $F_{EUV}$ that is 100 times larger than the typically
expected value. These high EUV flux levels can be realized for solar
systems located at distances $d \sim R/10$, i.e., in the central cores
of the clusters. Note that expected FUV fluxes ($G_0 \sim 3000$) can
evaporate the solar nebula at $r_d$ = 30 AU over a timescale of 15
Myr. Since giant planets are expected to form on somewhat shorter
timescales (3 -- 10 Myr, Lissauer \& Stevenson 2007), the solar nebula
is relatively safe. For more extreme fluxes with $G_0$ = 30,000, only
the outer 20 AU of the solar nebula can survive for 6 Myr. Thus,
survival of the early solar nebula, out to $r_d \sim 30$ AU, requires
that the FUV flux cannot exceed an intermediate value, i.e., 
$G_0 \le 10^4$. 

For completeness we note that Neptune and Uranus are ice giants,
rather than gas giant planets like Jupiter and Saturn. The relatively
low gas content in these bodies could imply that the early solar
nebula did in fact experience some photoevaporation near $r$ = 30 AU.
If this were the case, FUV flux levels near $G_0 \sim 10^4$ would be
required. Alternatively, these bodies could form over long timescales
such that less gas is present.

Observations of circumstellar disks in the Orion Trapezium Cluster, a
nearby region containing high mass stars and intense radiation fields,
indicate that the fraction of systems containing at least a MMSN
within $r = 60$ AU is $\sim12$ percent (Mann \& Williams 2009).  This
percentage is comparable to that in Taurus, a region with no high mass
stars and little background radiation. Taken together, these
observational results argue that cluster environments provide
relatively modest constraints on the mass available for planet
formation, a conclusion consistent with the theoretical considerations
outlined above.

\subsection{Ionization and Other Effects} 

Cluster environments provide important sources of ionization for
forming and newly formed solar systems.  These sources include
ionizing EUV radiation, X-rays, and cosmic rays. The distributions 
of EUV fluxes are described in Section 5.2 (see also Figure
\ref{fig:euvtime}). For larger clusters with membership $N \ge 100$,
the background cluster environment provides more ionizing photons to
the solar nebula than the early Sun itself (Adams \& Myers 2001). Most
of the external EUV photons are captured by the outer disk, whereas
most of the Solar EUV photons are intercepted by the inner disk; the
relative importance of the two ionization sources thus varies with 
radial position.

The case of X-ray radiation, with photon energy $h \nu \ge 0.1$ keV,
is similar: The typical X-ray luminosity from young stars falls in the
range $L_X \approx 10^{29} - 10^{32}$ erg s$^{-1}$ for stellar masses
in the range $M_\ast$ = 0.3 -- 7 $M_\odot$ (Preibisch et al. 2005).
For larger stars, $L_X \approx 10^{-6} L_\ast$.  With these
luminosities, X-rays provide fewer ionizing photons than the EUV
band. However, the EUV radiation is more easily absorbed, so that both
sources of radiation must generally be considered (see Gorti \&
Hollenbach 2009, Ercolano et al. 2009). 

Cosmic rays provide an important source of ionization, especially deep
within molecular clouds where star formation takes place and where UV
radiation can be shielded. Since supernovae are the source of cosmic
rays, and since they generally explode within or near molecular
clouds, cosmic ray fluxes can be enhanced relative to their standard
values in the interstellar medium. Further, since the clouds are
supported (at least in part) by magnetic fields, which act to retain
cosmic rays within the clouds, substantial enhancements are possible
(Fatuzzo et al. 2006). The short-lived radio isotopes (discussed in
the following section) also provide significant sources of ionization,
with $^{26}$Al being one of the most efficient (Umebayashi \& Nakano
2009).

Ionization levels are important for star formation and planet
formation.  In the early phases of star formation, molecular cloud
cores are supported, in part, by magnetic fields. Although the
relative importance of magnetic diffusion and turbulence is currently
under debate (compare Shu et al. 1987 with McKee \& Ostriker 2007),
loss of magnetic flux is necessary for stars to form. Increasing the
ionization increases the coupling between the field and the largely
neutral gas, and thereby decreases the ability of magnetic fields to
diffuse away. The density of ions $\rho_i$ in molecular clouds is
given by $\rho_i = {\cal C} \rho_n^{1/2}$, where $\rho_n$ is the
density of neutral atoms; the constant ${\cal C} \propto \zeta^{1/2}$,
where $\zeta$ is the flux of cosmic rays. The effective diffusion
constant $D$ for magnetic flux loss is then given by
\be
D = {v_A^2 \over \gamma_{in} {\cal C} \rho} \, 
\propto \zeta^{-1/2} \, , 
\ee
where $v_A$ is the Alfv{\'e}n speed and $\gamma_{in}$ is the drag
coefficient between ions and neutrals (Shu 1992). 

Ionizing radiation in clusters also influences disk accretion, which
is driven by an effective viscosity resulting from turbulence. This
turbulence, in turn, is thought to be driven by MHD effects such as
the magneto-rotational instability (MRI, Balbus \& Hawley 1991). The
presence of MRI and hence disk accretion requires that the ionization
fraction in the disk is high enough for the gas to be well-coupled to
the magnetic field.  The inner disk can be ionized by collisions, and
the outer disk can be ionized by cosmic rays. At intermediate radii,
however, the disk can have dead zones where ionization levels are too
low (Gammie 1996). Enhanced ionization in clusters thus acts to make
more of the disk support MRI. Thus, one consequence of the Sun forming
in a cluster is that disk accretion could be enhanced relative to the
rates it would have experienced in isolation.

Finally, we note that strong radiation fields can produce chemical
signatures in forming solar systems. Our own solar system displays an
oxygen isotopic anomaly that can be explained if the Sun formed in the
presence of intense FUV radiation fields. In one scenario, ultraviolet
radiation produces selective photodissociation of CO within the
collapsing protostellar envelope of the forming Sun (Lee et al. 2008);
in an alternate scenario, the isotope selective photodissociation
occurs at the surface of the early solar nebula (Lyons \& Young 2005).
Since a range of oxygen anomalies are possible, given current
measurements, the required FUV flux is not well determined.  Future
observations will provide much tighter constraints.

\section{CONSTRAINTS FROM NUCLEAR ENRICHMENT} 

\subsection{External Enrichment through Supernovae} 

As outlined in Section 2.7, meteoritic evidence implies that the early
solar nebula contained significant quantities of radioactive nuclei
with half-lives shorter than 10 Myr (see Table I). Supernovae provide
one possible source for these short-lived ratio isotopes. The idea of
a supernova explosion associated with the formation of the Solar
System has a long history. One of the first isotopes to be considered
was $^{26}$Al, which has a half-life of only 0.72 Myr.  To explain the
presence of $^{26}$Al, Cameron \& Truran (1977) suggested that
supernova ejecta containing the short-lived species could be
incorporated into the dense core that formed the Solar System (note
that asymptotic giant branch stars can also produce $^{26}$Al --- see
Section 6.5). This idea of external enrichment has been expanded upon
as additional nuclear species were discovered in meteorites (Table I).
In particular, the isotope $^{60}$Fe is extremely difficult to produce
through spallation reactions, but is naturally produced by stellar
nucleosynthesis. As a result, a number of authors have presented 
scenarios for supernova enrichment of the early solar nebula
(including Cameron et al. 1995, Boss \& Foster 1998, Goswami \&
Vanhala 2000, Looney et al. 2006, Williams \& Gaidos 2007, and many
others; see also references therein). Although a range of progenitor
masses $M_\ast$ are possible, and no mass scale produces perfect
abundances, these studies suggest that stars with $M_\ast \approx 25
M_\odot$ provide the best ensemble of short-lived radioactive nuclei.
This section outlines the basic requirements necessary for supernova
enrichment to take place, as well as the corresponding constraints on
the solar birth environment. Some of the difficulties faced by this
scenario are also discussed.

\begin{figure} 
\centerline{\epsscale{0.90} \plotone{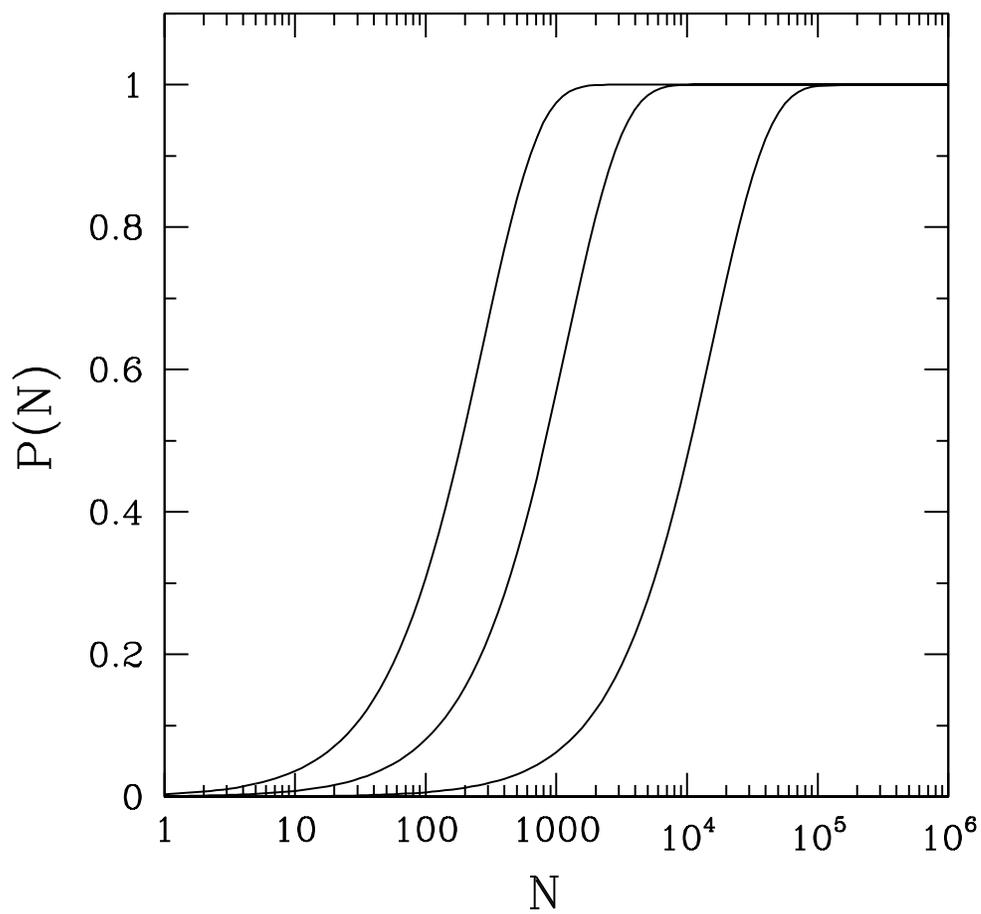}}
\caption{Probability for a cluster of membership size $N$ to 
produce a supernova progenitor of a given mass $M_\ast$ as a function
of $N$.  The three curves show the probability distributions for
progenitor masses of $M_\ast$ = 10, 25, and 75 $M_\odot$ (from left 
to right). } 
\label{fig:superp} 
\end{figure}

One needs a moderately large cluster to provide a supernova from a
sufficiently massive progenitor star. The stellar mass distribution of
equation (\ref{imf}) indicates that the probability that a star has
mass (in solar units) greater than a mass scale $m_0$ is given by
the expression 
\be
P (m \ge m_0) = \fone m_0^{-\gamma} \left[1 - 
\left( {m_0 \over m_\infty} \right)^{\gamma}\right] \, , 
\label{probone} 
\ee
where $m_\infty$ is the maximum stellar mass. For example, the
probability $P_{25}$ that a star has at least the benchmark progenitor
mass for supernovae enrichment, $m_0 = 25$, is given by $P_{25}
\approx 0.00084$ (where we have used standard values $\fone = 0.12$,
$\gamma$ = 1.5, and $m_\infty = 100$). The probability 
${\cal P}_N(m>m_0)$ that a system of $N$ stars contains at least one
star greater than mass $m_0$ is then given by 
\be
{\cal P}_N (m>m_0) = 1 - \left[ 1 - P(m>m_0) \right]^N \, . 
\label{probN} 
\ee
Throughout this analysis, we assume that high mass stars, specifically
those that can be progenitors of the supernova that enriched the early
solar nebula, are drawn at random from the IMF, and that this property
holds for all cluster sizes $N$. Although the largest stellar mass in
a system could in principle be correlated with cluster membership size
$N$, available data remain consistent with no such correlation,
especially for larger clusters (Maschberger \& Clarke 2008). The
resulting probability distributions for a cluster to produce a high
mass progenitor are shown in Figure \ref{fig:superp} for stellar
masses $m_0$ = 10, 25, and 75.

We define $N_{50}(m_0)$ to be the cluster membership size that is
required for the system to have a 50/50 chance of containing a star
with mass $m > m_0$.  In general, for mass $m_0$, this required
cluster size is given by
\be
N_{50} \approx {(\ln 2) \, m_0^\gamma \over \fone 
\left[ 1 - (m_0/m_\infty)^\gamma \right] } \, . 
\label{nfifty} 
\ee
The cluster size required to have a 50/50 chance of containing a 25
solar mass star is thus $N_{50} \approx 825$. For comparison, the
cluster size required to have a 50/50 chance of realizing a 75 solar
mass star is $N_{50} \approx 10,700$.  Provided that the progenitor
mass $m_0$ is not near the upper limit $m_\infty$, equation
(\ref{nfifty}) simplifies to the approximate form $N_{50} \approx 6 \,
m_0^{3/2}$, where we have inserted typical values for the remaining
parameters. Keep in mind that different progenitor masses have
different main sequence lifetimes, and that timing is also important 
for successful nuclear enrichment (see the discussion below).  

The distance $d$ from the supernova progenitor to the early solar
nebula must be close enough to provide the observed abundances of
radioactive isotopes.  The nominal distance from the supernova
explosion to the solar nebula can be estimated by requiring the mass
fraction $X_j$ of a given nuclear species to be large enough. This
fraction is given by
\be
X_j  = f_j {M_j \over M_d} {\pi r_d^2 \over 4 \pi d^2} \, , 
\label{sndistance}
\ee
where $M_d$ is the mass of the solar nebula at the time of enrichment,
$f_j$ is the fraction of the material that is absorbed by the nebula,
and $r_d$ is its radius (Looney et al. 2006, Ouellette et al. 2007). 
For example, calculations of radioactive yields by Type II supernovae
indicate that such explosions produce $M_{\rm Fe}$ = 2.4 -- 16 $\times
10^{-5}$ $M_\odot$ of $^{60}$Fe, where the value depends on the
progenitor mass (Rauscher et al. 2002).  Using this result and the mass
fraction $X_{\rm Fe}$ for $^{60}$Fe (see Table I), and parameters of
the minimum mass solar nebula ($M_d$ = 0.05 $M_\odot$, $r_d$ = 30 AU),
we find the required distance to fall in the range $d \approx$ 0.1 --
0.3 pc.  This estimate assumes that all of the material is accreted by
the nebula so that $f_j$ = 1; the inclusion of an efficiency factor
$f_j \ne 1$ implies an even closer distance. This estimate also
assumes that the nebula is facing into the blast; significant
inclination angles will also reduce the estimated distance. On the
other hand, non-uniform ejecta (clumps) can lead to greater yields and
allow for a larger distance.  Nonetheless, in order of magnitude, the
required distance for sufficient enrichment is $d \sim 0.2$ pc.

On the other hand, a minimum mass solar nebula will be stripped by a
supernova blast wave if it lies too close to the explosion. This
minimum distance is also estimated to be about 0.2 pc (Chevalier
2000), as outlined below. As a result, there is some tension between
the requirement that the supernova is close enough to produce
sufficiently high yields of the radio isotopes and yet far enough that
the early solar nebula survives.

\begin{figure} 
\centerline{\epsscale{0.90} \plotone{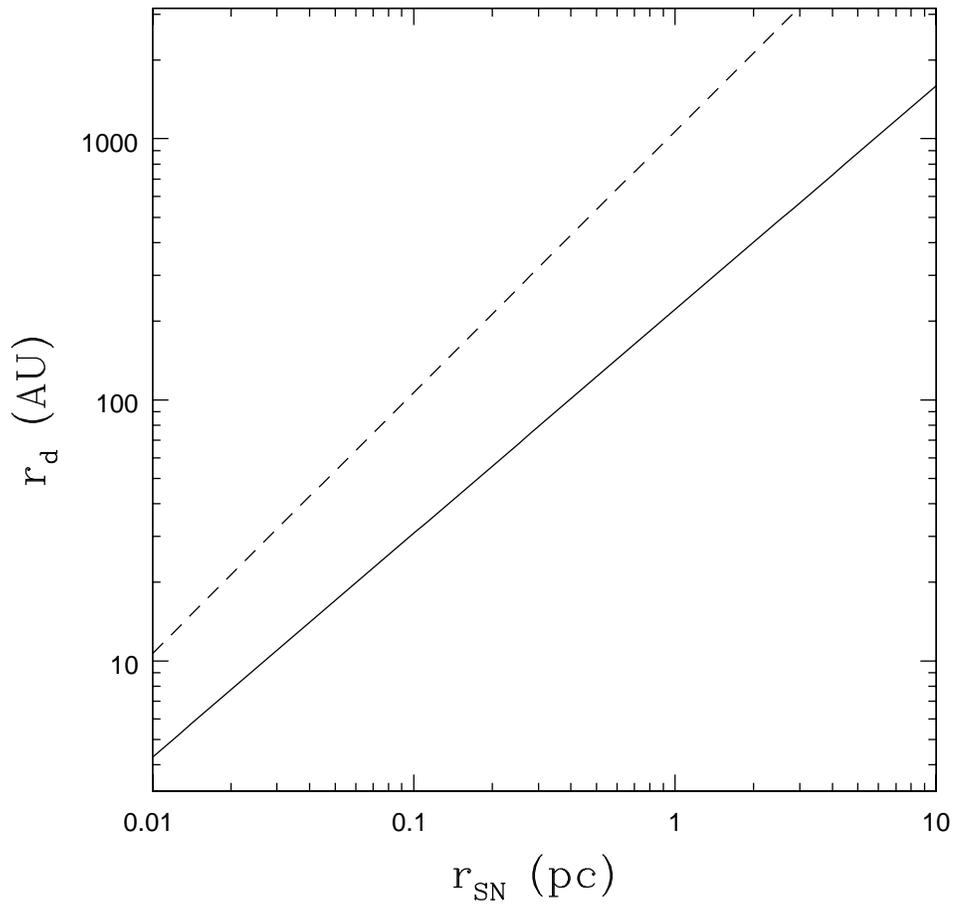}}
\caption{Disk radius $r_d$ that can survive a supernova blast wave 
as a function of the distance $r_{SN}$ from the star/disk system to 
the explosion. The solid line shows the disk radius that can survive 
ram pressure stripping; the dashed line shows the disk radius that 
can survive momentum stripping. } 
\label{fig:strip} 
\end{figure}

The early solar nebula can be truncated by a supernova explosion in
two ways. When the ram pressure $P_{ram} = (\rho v^2)_{SN}$ of the
supernova flow exceeds the force per unit area with which the Sun
holds onto the nebular gas, the material is subject to stripping.
This second pressure scale is given roughly by $P_\odot \sim G M_\ast
\Sigma / r_d^2$, where $\Sigma$ is the disk surface density and $r_d$
is the radial location within the disk. For supernovae, the ram
pressure $(\rho v^2)_{SN} \approx A_{SN} E_{SN} / r^3$, where
$E_{SN} \approx 10^{51}$ erg, $r$ is the distance to the explosion,
and the dimensionless parameter $A_{SN}$ is of order unity
(Chevalier 2000). The condition for ram pressure stripping thus 
takes the form 
\be
A_{SN} {E_{SN} \over r^2} = {G M_\ast \over r_d^2} \Sigma(r_d) \, . 
\ee
Similarly, the early solar nebula can be destroyed by momentum
transfer if the momentum per unit area imparted by the supernova blast
wave exceeds the corresponding scale of the star/disk system. This
criterion for momentum stripping can be written in the form 
\be
{M_{ej} v_{SN} \over 4 \pi r^2 } = 
\left( {2 G M_\ast \over r_d} \right)^{1/2} \Sigma (r_d) \, ,
\ee 
where $M_{ej} \approx 1 M_\odot$ is the mass of the ejecta. The
maximum radii of the solar nebula that can survive these two types of
stripping processes are shown in Figure \ref{fig:strip}. The curves in
this figures are calculated for a minimum mass solar nebula, and for 
the values of $M_{ej}$ and $E_{SN}$ given above (note that $v_{SN}^2$ 
= $2 E_{SN}/M_{ej}$).  In general, ram pressure stripping is more
destructive than momentum stripping. These results indicate that the
supernova explosion must be farther away than $r = d \sim 0.1$ pc in
order for the solar nebula, with outer radius $r_d \approx 30$ AU, to
survive intact. Note that recent numerical studies indicate that the
solar nebula can survive at even closer distances (Ouellette et
al. 2007).

The above considerations show that the early solar nebula must be
close enough to the supernova ($d \le d_2 \approx 0.3$ pc) to receive
enough nuclear enrichment, and far enough away ($d \ge d_1 \approx
0.1$ pc) to survive the experience.  It is significant that an
intermediate range of radii allows for both conditions to be met. 
However, the probability of the solar nebula residing in this range of
radii is relatively low: For example, the radial probability
distribution from equation (\ref{dpdr}) implies that the probability
$P_d$ of a star finding itself in the radial range 0.1 pc $\le d \le$
0.3 pc is given by $P_d = 0.08 ({\rm pc})^2/R^2 \approx 0.02$. The
latter numerical value assumes that the cluster radius $R$ = 2 pc, a
typical value for a moderately large cluster (see equation
[\ref{radlaw}]). One can generalize this result to include cluster
density profiles of the form $n_\ast \propto r^{-p}$ with outer radii
given by equation (\ref{radlaw}):
\be 
P_d = \left[ (d_2/R_0)^{3-p} - (d_1/R_0)^{3-p} \right] 
\left( {N_0 \over N} \right)^{\alpha (3-p)} \, ,
\label{generalrad} 
\ee
where $R_0$ = 1 pc and $N_0$ = 300; we expect the indices to fall
within the ranges $1 \le p \le 2$ and $1/4 \le \alpha \le 1/2$.
The typical value is thus $P_d \sim$ few percent. 

The lifetimes of potential progenitors provide another strong
constraint on the supernova enrichment hypothesis. As outlined above,
the radio isotope yields work best for progenitor masses $M_\ast$
$\approx 25 M_\odot$.  Stars with this initial mass spend about 6.7
Myr in their main-sequence hydrogen burning phase, and a total of
$\sim7.54$ Myr before core collapse (Woosley et al. 2002). These timescales
decrease with increasing starting mass: For comparison, stars with
initial mass 15 $M_\odot$ (75 $M_\odot$) spend 11.1 Myr (3.16 Myr) on
the main sequence and a total of 12.1 Myr (3.64 Myr) before exploding
as supernovae. Given that age spreads in young embedded clusters are
relatively small, only about 1 -- 2 Myr (Hillenbrand 1997), and that
most clusters are dispersed in $\sim10$ Myr or less (Allen et al. 
2007), these progenitor lifetimes are somewhat long for comfort.
Nonetheless, a possible solution exists, provided that the massive
star forms first and the Sun forms several Myr later. In this case,
however, the forming Solar System could be less likely to reside
within the required nuclear enrichment zone ($d$ = 0.1 -- 0.3 pc) due
to radiative disruption from the pre-existing progenitor.

In addition to the lifetime issue, the necessity of having a large
progenitor star to explain nuclear enrichment introduces another
strong constraint. The massive progenitor star will produce large
amounts of EUV and FUV radiation, which can readily evaporate the
early solar nebula (see Section 5, Gounelle \& Meibom 2008). If the
Sun spends too much time close to the progenitor, before the
explosion, the solar nebula could be compromised. For example,
consider a cluster with $N$ = 1000 members and let its UV luminosity
be given by equation (\ref{luvmean}); if the solar system orbits at a
distance from the center with $d$ = 0.3 pc, the expected FUV flux has
a value $G_0 \approx 60,000$. This radiation field can readily
evaporate the solar nebula, and would remove all gaseous material
beyond $\sim10$ AU over a time span of 10 Myr (see Figure
\ref{fig:fuvtime}).  As a result, there is significant tension between
the radiation fields produced by the progenitor and the relatively
close proximity required for successful nuclear enrichment. Of course,
the Sun could spend much of its time (during the main-sequence phase of
the progenitor) in the outer parts of the cluster, where the radiation
fields are low.  The Sun would then have to enter the central part of
the cluster just before the supernova explosion. This timing of events
is possible, since most stars spend most of their time outside the
inner enrichment zone of a cluster. However, this requirement lowers
the odds for supernova enrichment (see also Williams \& Gaidos 2007).

\subsection{Triggering of Solar System Formation} 

Two versions of the external enrichment scenario have been proposed.
In the first and simplest case, the supernova enriches the Solar
System after it has already begun formation. Here, the supernova
ejecta are intercepted by the early solar nebula, as discussed above.
In the second version, the supernova that enriches the Solar System
also triggers the initial collapse of the molecular cloud core that
gives rise to the Sun (Cameron \& Truran 1977, Vanhala \& Boss 2002).
In this case, the supernova ejecta are incorporated into the molecular
cloud material that subsequently collapses to form the Solar System.

Current observational data suggest that the latter, directly triggered
scenario is less likely than the alternative. The cluster environments
found in the solar neighborhood (Lada \& Lada 2003, Porras et al. 
2003, Allen et al. 2007) are generally not influenced by supernovae:
Although the data remain incomplete, the spread in ages is these
clusters is small, typically less than $\sim1$ Myr (Allen et al.
2007); since this timescale is shorter than the main sequence lifetime
of supernova progenitors, none of the stars in these cluster systems
were triggered to collapse by supernovae exploding within the same
cluster. In a similar vein, the statistics of these clusters suggest
that the gas is removed, and hence that star formation shuts down,
after only 3 -- 5 Myr (e.g., Gutermuth et al. 2009, Lada \& Lada
2003). This timescale is (again) shorter than the main sequence
lifetime of most supernova progenitors (Section 6.1), so that no gas
is expected to be left in the cluster when stellar explosions
eventually occur.

For completeness we note that although supernova triggering does not
seem to take place within the embedded clusters observed in the solar
neighborhood, triggering mechanisms are not completely ruled out. In
particular, collapse can be induced, or at least helped along, by
ionization fronts driven by massive stars (for further discussion of
latter this issue, see Hester \& Desch 2005, Snider et al. 2009). In
addition, supernovae do in fact shape star formation environments, but
their actions take place over time scales ($\sim10$ Myr) and length
scales (many pc) that are larger than those appropriate for star
formation within a single cluster --- there is an impedance mismatch 
between supernova triggering and cluster scales. 

\subsection{Internal Enrichment through Irradiation} 

Some fraction of the short-lived radio isotopes can be produced by
internal processes, i.e., by the forming Sun itself. Note that if all
of the observed nuclear species could be produced internally, then
nuclear enrichment would not require an external supernova or other
source of short-lived isotopes. In that case, the birth environment of
the Sun would be decoupled from nuclear considerations. In spite of
the progress made over the last decade (Lee et al. 1998, Shang et al.
2000, Shu et al. 2001), it remains difficult for internal processes to
explain the entire ensemble of isotopes, so that some supernova
enrichment is still (apparently) required.  On the other hand,
self-enrichment may be required to explain the presence of some
nuclear species and is thus likely to contribute to the supply of
other isotopes. In this manner, the requirements placed on external
sources are made less restrictive. 

The leading picture for internal enrichment is generally called the
X-wind model (Shu et al. 2004). During the protostellar collapse phase
of the early Sun, a wind emerges from the inner portion of the rapidly
rotating solar nebula. The Sun and its circumstellar disk are coupled
by strong magnetic fields anchored within the star; the disk is also
influenced by additional magnetic fields that are dragged in from the
original cloud. These fields drive a powerful stellar wind through
magnetic/centrifugal effects, and the wind collimates into a narrow
jet with a molecular outflow. As rocks enter the launching region of
the wind, they are lifted up, heated and irradiated by both photons
and energetic particles from the stellar surface, and then thrown
outward.  Small bodies can be carried out of the Solar System along
with the wind, whereas much larger bodies move relatively little and
fall near their launching locations. The rocks of intermediate mass,
those with sizes in the range 0.2 -- 2 mm, fall out of the wind at
radii comparable to the present day asteroid belt, where they can be
incorporated into meteoritic material (Shang et al. 2000).

This X-wind environment provides both heating of the rocky material
that it processes and the production of radioactive nuclear species.
Although both processes are important for understanding the properties
of our Solar System, the latter has a more direct bearing on the
possible birth place of the Sun. In this setting, radio isotopes are
produced by energetic particles --- essentially cosmic rays --- that
are released by energetic protostellar flares near the surface of the
star.  These high energy cosmic rays induce spallation reactions,
which, in turn, can synthesize some of the short-lived radioactive
isotopes that are thought to be present in the early solar nebula.

In the present day Sun, gradual flares dominate the production of the
cosmic rays that can leave the Solar System because these flares
operate on open magnetic field lines. Impulsive flares on closed field
lines provide more energetic particle displays.  In the early Sun,
these impulsive flares are thought to dominate (see the discussion of
Shang et al. 2000) and create a prolific source of $^3$He nuclei, in
addition to alpha particles and energetic protons.  Large fluxes of
$^3$He can readily interact with stable isotopes of intermediate mass
nuclei and thereby produce short-lived radioactive species such as
$^{26}$Al, $^{36}$Cl, and $^{41}$Ca (Gounelle et al. 2001, 2006).
Note that these species are also produced by supernovae (Woosley et
al. 2002, Wadhwa et al. 2007).  More importantly, however, these
spallation reactions can produce the light isotopes $^7$Be and
$^{10}$Be, species that are not produced via stellar nucleosynthesis,
and hence are not explained by supernova enrichment. The detection of
$^{10}$Be in an Allende inclusion (McKeegan et al. 2000) thus argues
for internal irradiation (Gounelle et al. 2001). However, this latter
conclusion assumes that $^{10}$Be cannot be produce via spallation
from Galactic cosmic rays, and this claim has been disputed (Desch et
al. 2004).

It is important to keep in mind that $^{60}$Fe cannot be produced
through internal irradiation and hence requires an external source.
Current data from meteorites can thus be interpreted to suggest that
both external and internal enrichment mechanisms are required.
Specifically, the local enrichment scenario has difficulty producing
$^{60}$Fe, which argues for a supernova origin; in contrast, $^{10}$Be
is produced only by spallation processes, which argues for internal
enrichment.

\subsection{Distributed Supernova Enrichment} 

Although $^{60}$Fe must be produced through stellar nucleosynthesis,
the enrichment of the early solar nebula does not necessarily require
a single supernova source. The generally accepted abundance of
$^{60}$Fe is quoted in Table I, but measurements of this quantity
remain uncertain.  For example, this value could actually be an upper
limit, with the true abundance lower by a factor of $\sim3$ (see
Gounelle \& Meibom 2008, and references therein). In light of this
possible revision, one recent model suggests that a collection of
supernovae, all taking place within the same molecular cloud over at
timescale of 10 -- 20 Myr, could account for the observed iron
abundances (Gounelle et al. 2009).  

Since one supernova is a low probability event, one might worry that a
collection of supernovae would be rare: Consider a $10^6$ $M_\odot$
molecular cloud. With typical star formation efficiencies, the cloud
can produce perhaps $3 \times 10^4$ stars over the 10 -- 20 Myr time
span of interest. Given the IMF presented in Section 2.1, this
population of stars is expected to produce $\sim140$ supernova
progenitors (with masses $m \ge 8$).  However, this potential
$\sim$100-fold increase in supernova numbers (compared with enrichment
by a single event) must be balanced against the increased source
distances and the loss of material through radioactive decay (the
half-life of $^{60}$Fe is only $\sim2$ Myr). On the other hand, the
target area for capture is greatly increased, and the fact that
injection takes place in a diffuse phase helps efficiency.  In
addition, the point-to-point variation in the amount of $^{60}$Fe
produced could be substantial, so that Sun could form in a region with
a positive fluctuation in $^{60}$Fe abundance.  Finally, we note that
recent measurements (Rugel et al. 2009) suggest that the half-life for
$^{60}$Fe could be longer (about 2.5 Myr) than the accepted value (1.5
Myr), which would make this distributed enrichment scenario easier to
realize.

Recent work (Connelly et al. 2008) measures the age of the lead system
$^{207}$Pb--$^{206}$Pb in the chondrules of Allende, and finds an age
of 4565.45 $\pm$ 0.45 Myr. This age is younger than the standard age
for calcium-aluminum-rich inclusions (CAIs), and agrees with the ages
found for the $^{26}$Al--$^{26}$Mg system. Further, these isotopes are
inferred to be distributed homogeneously throughout the solar nebula
(Villeneuve et al. 2009).  Since the lead isotopes arise from stellar
nucleosynthesis, this finding offers support for the theory of
supernova enrichment of $^{26}$Al. On the other hand, if these nuclear
species were injected by a single supernova, the $^{26}$Al might not
have been so evenly distributed across the early solar nebula (whereas
internal irradiation and/or distributed supernovae could account for
this homogeneity).

\subsection{Other Enrichment Scenarios} 

Although the current consensus holds that short-lived radioactive
nuclei are provided to the early solar system by either supernovae or
internal sources --- or perhaps both --- one should keep in mind that
other possibilities exist. For example, thermally pulsating asymptotic
giant branch (AGB) stars have been suggested as an enrichment source
(Busso et al. 1999, 2003). Since these stars only provide radioactive
nuclei at the end of their lives, and since their main-sequence
lifetimes are long, the probability that such a source would be
associated with a molecular cloud is relatively low (Kastner \& Myers 1994).
The abundance yields work best for AGB stars of intermediate mass $M_\ast$
= $3 - 5 M_\odot$ (Wasserburg et al. 2006), and such stars have
main-sequence lifetimes in the range 70 -- 200 Myr. These timescales
are thus longer than the lifetimes of molecular clouds, and hence the
timing problem for seeding the early solar nebula is more severe for
models using AGB stars than for models using supernovae. Nonetheless, 
enrichment scenarios using asymptotic giant branch stars have been
constructed (Trigo-Rodriguez et al. 2008) and should be considered further.

Wolf-Rayet stars represent another external source of radioactive
nuclei for the early solar nebula. This enrichment mechanism still
requires a massive star, here with progenitor mass $M_\ast > 60$
$M_\odot$ (Arnould et al. 1997), but without the explosion. Since such
massive stars are much rarer than those required for supernova
enrichment, this scenario is somewhat less probable. In this case we
make a quantitative estimate using equation (\ref{probone}): The
chances of finding a 60 solar mass star are smaller than the chances
of finding a 25 solar mass star by a factor of $\sim5.5$. On the other
hand, the stellar lifetime is shorter ($\sim5$ Myr), which helps with
timing issues.  Notice also that Wolf-Rayet winds could supply some
fraction of the observed short-lived radio isotopes, $^{26}$Al for
example (Gaidos et al. 2009), in addition to enrichment from other
sources (supernovae and/or internal irradiation).

Finally, for completeness, we reiterate that short-lived radio
isotopes not only provide constraints on the environment in which the
Sun formed, but also provide heating sources (Hester \& Desch
2005). For example, the decay of $^{26}$Al provides a substantial
supply of energy for the differentiation of planetesimals (Grimm \&
McSween 1993). On the other hand, the X-wind mechanism, which can
provide internal nuclear enrichment, also acts as a heating source
(Lee et al. 1998).  Elaborate models have been constructed (Shang et
al. 2000, Shu et al. 2001) to account for the signatures of heating
found in both chondrules and CAIs.

\section{SUMMARY} 

\subsection{Overview of Results} 

This review has outlined a number of constraints on the star formation
process that produced our Solar System. The first set of issues
concerns the general properties of our Sun and planets, and provides
us with an assessment of whether Sun-like systems are common or rare.
The Sun is a relatively massive star (Section 2.1) and stars this
large are expected to form 12 percent of the time. The Sun is a single
star, which occurs about 30 percent of the time for solar-mass stars,
but more often for the general stellar population (Section 2.2).  The
Sun has relatively high metallicity (Section 2.3), placing our solar
system in the top 25 percent.  Our Solar System has successfully made
giant planets, a feat that is accomplished by about 20 percent of
solar-type stars (Section 2.4).  The outer edge of the solar system
(Section 2.5) indicates that the early solar nebula extended out to 30
-- 50 AU, a size that is typical among star/disk systems observed
today.  All of these required features of the Solar System are thus
relatively common.

It is important to keep in mind that the chances of a solar system
realizing all of the properties outlined above are far from
guaranteed.  For example, we can write down an analog of the Drake
equation to assess the combined probability ${\cal P}_\odot$ for a
given solar system to have the above characteristics, 
\be
{\cal P}_\odot = \fone {\cal F}_Z {\cal F}_B {\cal F}_P 
\dots \le 0.0018 \, , 
\label{drake} 
\ee
where the factors correspond to the probabilities for a solar system
to have at least a one solar mass star, at least solar metallicity, no
binary companion, form giant planets, and so on. The numerical value
on the right hand side of the equation provides an upper limit to the
probability, provided that the factors are statistically independent. 
Although this probably is low (less than one percent), one should not
conclude that that solar systems like ours are rare or unusual. The
necessity of having a large number of relatively common properties
results in a low probability for the combination to occur.  However,
with $\sim$100 billion stars in the Galaxy, the probability would have
to be much lower for our Solar System to be considered unusual (see
also the discussion of Gustafsson 2008).

The above considerations are (mostly) independent of the particular
birth environment of the Sun. Additional properties of the Solar
System allow us to constrain the properties of the solar birth
cluster. We first consider dynamical constraints (Section 4): The
observed planetary orbits indicate that no passing stars have made
disruptive close encounters with the Solar System after the giant
planets were produced, where the closest possible approach is about
225 AU. The early solar nebula extended out to approximately 30 AU,
which indicates that no passing star came closer than about 100 AU at
earlier epochs. On the other hand, the observed orbital elements of
the dwarf planet Sedna can be understood if a close encounter did take
place, where the required distance of closest approach $b$ = 400 --
800 AU. Stellar encounters at much closer distances tend to produce
too many Sedna-like objects, so that closer encounters are unlikely.
All of these system properties can thus be understood if the early
solar system experienced an encounter with a distance of closest
approach $b \sim 400$ AU. This requirement, in turn, constrains the
stellar density $n_\ast$ of the birth environment and the Solar
System's residence time $t$ in that region (Section 4.4) so that
$\langle n_\ast t \rangle \approx 80,000$ pc$^{-3}$ Myr. Since the
typical mean stellar density is only of order $n_\ast \sim 100$
pc$^{-3}$, the Solar System must live within its birth cluster a
relatively long time (perhaps a few hundred million years), or live in
a somewhat higher density environment, in order to experience the
required close encounter. Only relatively large bound clusters, those
with $N \ge 1000$, are expected to live that long (Binney \& Tremaine
1987, Kroupa et al. 2001, Lamers et al. 2005). 

Clusters also provide external radiation fields that affect the
forming Solar System, primarily by evaporating the early solar nebula
(Section 5).  Although cluster environments provide both EUV and FUV
radiation, the latter tends to dominate the photoevaporation of disks
(Figures \ref{fig:euvtime} and \ref{fig:fuvtime}). An external FUV
flux with $G_0$ = 3000 will evaporate the outer part of the solar
nebula (beyond about 36 AU) over 10 Myr, the typical timescale for
disks to survive and for giant planets to form. Since the nebula must
have retained its gas within Neptune's orbit (Section 2), the FUV flux
cannot be much larger than this benchmark value. Although the solar
nebula could have formed with an outer radius $r_d \sim 30$ AU, the
disk could also have been larger, and we can assess what type of
cluster is necessary to provide an explanation for this outer radius
through photoevaporation: The required radiation field is about $G_0
\approx 10^4$ (Figure \ref{fig:fuvtime}), a value that is cleanly
beyond the peak of the distribution for young clusters in the solar
neighborhood (Figure \ref{fig:fuvdist}). As a result, the birth
cluster must be relatively large, say, with $N \ge 1000$. Note that
the Solar System could have been born within an even larger cluster,
provided that it (primarily) resided at large radii until the gas
giant planets were produced.

The next set of constraints on the solar birth cluster arises from the
required presence of short-lived radio isotopes (Section 6). As
outlined above, some fraction of short-lived radioactive species must
have an origin from stellar nucleosynthesis, so that some enrichment
from a nearby supernova is indicated. The preferred starting mass for
the exploding star is $M_\ast \sim 25 M_\odot$.  In addition to
providing a good mix of short-lived radioactive isotopes, this mass
scale is suggested by cluster considerations: Smaller stars spend too
much time on the main-sequence and make the timing issues more
problematic. Larger stars are exceedingly rare, which pushes the
required cluster size to larger $N$ (see equation [\ref{nfifty}]),
which in turn leads to greater disruption of the solar nebula.  Given
the need for a large progenitor mass, and the rarity of massive stars,
supernova enrichment requires a large solar birth cluster with $N \ge
1000$ (see Figure \ref{fig:superp}). In addition, at the time of the
supernova explosion, the solar nebula must be close enough to capture
a sufficient amount of ejecta (equation [\ref{sndistance}]) and yet
far enough away to survive the blast (Figure \ref{fig:strip}).  This
compromise implies that the Solar System had to be roughly 0.2 pc from
the explosion, which most likely occurred near the cluster center
(Section 6.1). Keep in mind that these constraints can be alleviated
if some of the observed nuclear enrichment arises from internal sources
(Section 6.3) and/or distributed supernovae (Section 6.4).

These constraints are summarized in Table II, which lists the effects
outlined above, their implications, and the fraction of forming solar
systems that are expected to meet each requirement. These fractions
are approximate and are thus subject to future revision. The top four
entries correspond to Solar System properties that are largely
independent of the birth environment, whereas the bottom entries
depend on the cluster properties.  To assess the odds of a solar
system being born within a cluster of membership size $N$, we assume
that the probability is uniform-logarithmically distributed in $N$
(Section 3.1), and then use either dynamical considerations (Section
4) or supernova probability distributions (Section 6.1 and Figure
\ref{fig:superp}).  To assess the odds of the Solar System residing at
a given radial location, we use the $dP/dr$ distributions discussed in
Section 3.2 with density profile $n_\ast \propto r^{-2}$ (the form
expected for more evolved clusters) and the cluster radius law $R
\propto N^{1/3}$.  Note that the probability of surviving the
supernova ($d \ge 0.1$ pc) is not independent of the probability of
receiving enough ejecta ($d \le 0.3$ pc), so that the joint
probability is not their product (see equation [\ref{generalrad}]).
The odds of a solar system experiencing a given FUV radiation field is
determined from the flux distribution shown in Figure \ref{fig:fuvdist}. 
Notice that each individual constraint on the early solar system is
likely to be satisfied with reasonably high probability.  In this
sense, our particular star and planetary system are not rare or
unusual. However, as discussed above, the likelihood of a solar system
meeting all of these conditions is much lower (less than 1 percent).

\noindent
$\,$

\centerline{\bf Table II: Summary of Constraints} 

\bigskip
\begin{center} 
\begin{tabular}{lcc} 
\hline
\hline 
Solar System Property & Implication & Fraction \\ 
\hline
\hline
Mass of Sun & $M_\ast \ge 1 M_\odot$ & 0.12 \\ 
Solar Metallicity & $Z \ge Z_\odot$ & 0.25 \\ 
Single Star & (not binary) & 0.30 \\
Giant Planets & (successfully formed) & 0.20 \\
\hline  
Ordered Planetary Orbits & $N \le 10^4$ & 0.67 \\ 
Supernova Enrichment & $N \ge 10^3$ & 0.50 \\ 
Sedna-Producing Encounter & $10^3 \le N \le 10^4$ & 0.16 \\ 
Sufficient Supernova Ejecta & $d \le 0.3$ pc & 0.14 \\ 
Solar Nebula Survives Supernova & $d \ge 0.1$ pc & 0.95 \\ 
Supernova Ejecta and Survival & 0.1 pc $\le d \le$ 0.3 pc & 0.09 \\ 
FUV Radiation Affects Solar Nebula & $G_0 \ge 2000$ & 0.50 \\ 
Solar Nebula Survives Radiation & $G_0 \le 10^4$ & 0.80 \\ 
\hline  
\hline 
\end{tabular} 
\end{center}

\subsection{Scenarios for the Solar Birth Aggregate} 

Although the birth environment of the Solar System is significantly
constrained, one can find working scenarios that meet all of the
observational requirements. As a starting point, this section explores
the case where the Sun formed within a moderately large cluster with
$N = 10^3 - 10^4$, and outlines the cluster properties and other
considerations that are necessary to explain the observed system
properties. This solution is not unique, and it contains significant
shortcomings. Both of these issues are discussed below, as well as
some alternatives. In spite of the uncertainties, this discussion
demonstrates that a working scenario can be found, and illustrates the
highly constrained nature of the problem.

Most stars form in clusters of some size $N$. External enrichment of
short-lived radioisotopes suggests a cluster with at least $N \ge$
1000 in order to have a reasonable chance of producing a 25 $M_\odot$
star (the preferred progenitor mass for nuclear enrichment). However,
stellar models show that nucleosynthesis cannot provide the early
solar nebula with the full inventory of short-lived radio isotopes
(including $^{26}$Al, $^{36}$Cl, $^{41}$Ca, and $^{60}$Fe). For
example, models with a $M_\ast = 25 M_\odot$ progenitor can provide
the correct abundances of $^{26}$Al, $^{41}$Ca, and $^{60}$Fe, but
fail to produce the abundance of $^{36}$Cl by a factor of $\sim100$
(Meyer 2005). This discrepancy thus argues for a dual origin of the
radionuclides of intermediate atomic number (e.g., Gounelle et
al. 2006). This point of view is bolstered by the discovery of the
light isotopes $^7$Be and $^{10}$Be, which must be produced by
spallation rather than nucleosynthesis in stars. In this scenario,
local irradiation models produce the light isotopes, while a supernova
produces the proper abundance of $^{60}$Fe. Both stellar
nucleosynthesis and local irradiation models can deliver $^{26}$Al,
$^{36}$Cl, $^{41}$Ca. For completeness, note that a faint supernova
with mixing and fallback can also help explain the initial abundance
patterns of the short-lived radio isotopes (Takigawa et al. 2008). In
addition, distributed supernovae can produce $^{60}$Fe (Gounelle \&
Meibom 2008), and the $^{60}$Fe half-life could be longer (Rugel et
al. 2009), which would alleviate the some of the constraints implied
by the observed iron adundances.

Clusters in this membership size range $N = 10^3 - 10^4$ produce
strong radiation fields and significant probabilities for close
encounters. Although both of these effects can potentially cause
disruption, the early Solar System stands a good chance of surviving
unscathed.  On the other hand, the requirement of an encounter with 
$b \sim 400$ AU to explain the observed orbit of Sedna also argues for
a birth cluster in the approximate range $N$ = $10^3 - 10^4$.  Close
encounters are relatively rare in clusters with $N \le 1000$, in part
because of their short lifetimes; close encounters become more likely
with increasing $N$, so that bound clusters with $N \ge 10^4$ are
sufficiently long-lived that severe disruption becomes likely. Note
that the need for a Sedna-producing encounter implies that the Sun
formed within a gravitationally bound cluster, which occurs about 10
percent of the time.  The radiation from a cluster in this size range
will provide some disk evaporation, but the early solar nebula can
survive (for radii $r \le 30$ AU) as long as the Solar System does not
reside in the core of the cluster.  However, when the supernova
explosion ignites, the solar nebula must be only about 0.2 pc away,
which places it relatively near the core, certainly in the inner
portion of the cluster. These location restrictions lower the odds of
the Solar System achieving successful enrichment (see equation 
[\ref{generalrad}]). 

The timing requirements provide additional tight constraints on the
supernova enrichment hypothesis: The Sun and the progenitor are most
likely to form at nearly the same epoch, consistent with observations
of narrow age spreads in embedded young clusters. The progenitor can
burn through its fuel and then explode $\sim7.5$ Myr later. At this
time, the solar nebula could still have enough mass to capture the
required ejecta (with the observed disk ``half-life'' of $\sim3$ Myr,
about 20 percent of disks live this long). Nonetheless, this picture 
works better if the Solar System forms somewhat later, with a time
offset of a few to several Myr. For completeness we note that recent
measurements (Bizzarro et al. 2007) suggest that the oldest
planetesimals formed in the absence of $^{60}$Fe, with a $\sim1$ Myr
time delay between the oldest bodies and those that contain $^{60}$Fe;
however, although this data work in favor of the late enrichment
picture, subsequent work indicates that iron meteorites have the same
isotopic composition as the Earth, and hence does not find evidence
for this time difference (Dauphas et al.  2008). Notice that if the
progenitor has an even larger mass, its pre-explosion lifetime is
shorter, but the probability of a given cluster producing such a large
star decreases.  Relatively soon after the explosion, giant planet
formation is complete, but the Solar System remains in its birth
cluster. After this time, at an age of about 10 Myr, radiation from
the background cluster has a less destructive influence. The Solar
System must stay inside the cluster long enough for a close encounter
to provide Sedna with its observed orbital elements and perhaps to
help truncate the outer edge of the Kuiper belt. After this close
encounter, most likely when the cluster age is 10 -- 100 Myr, the
Solar System leaves its birth cluster with minimal additional
disruption.

\begin{figure} 
\centerline{\epsscale{0.90} \plotone{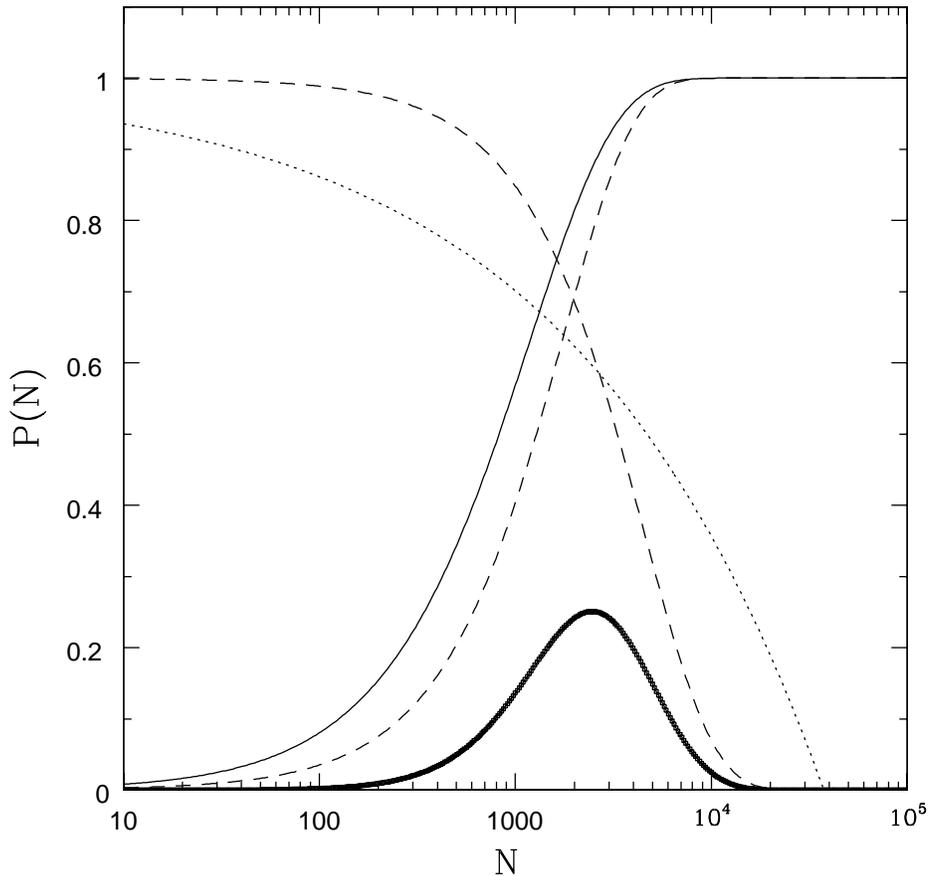}}
\caption{Probability of the solar birth cluster meeting several 
contraints as a function of stellar membership size $N$. Solid curve
shows the probability of the cluster producing a supernova with
progenitor mass $M_\ast \ge 25 M_\odot$. The dashed curves show the
probability of a close encounter with $b \le 400$ AU (to produce
Sedna), but no encounters with $b \le 225$ AU (to preserve planetary
orbits).  Dotted curve shows the probability of the FUV radiation
field having $G_0 \le 10^4$. The heavy bell-shaped curve shows the
joint probability distribution, for which $\langle N \rangle$ = 
$4300 \pm 2800$. These constraints are necessary but not sufficient:
For successful nuclear enrichment, the Solar System must also be
located the proper distance from the supernova and satisfy the
corresponding timing constraints (see text). }
\label{fig:joint} 
\end{figure}

This picture of the solar birth aggregate is specified further by the
probability distributions shown in Figure \ref{fig:joint}. The solid
curve shows the probability of a cluster producing a $M_\ast$ = 25
$M_\odot$ star as a function of stellar membership size $N$. The two
dashed curves show the probability that the solar nebula experiences
an encounter close enough to explain Sedna ($b \sim 400$ AU) and does
not experience an encounter so close that the orbital elements of the
planets are significantly changed ($b \sim 225$ AU). To produce these
curves, we have used the cluster properties outlined in Section 3.2,
including the lifetime estimate for bound clusters (from Lamers et al.
2005).  The dotted curve shows the probability for the Solar System to
experience an FUV radiation field less intense than $G_0$ = $10^4$;
larger values would evaporate too much gas from the region of the disk
that produces giant planets.  This condition requires the Solar System
to reside in the outer part of the cluster, where we have used the
radial probability distribution of equation (\ref{dpdr}). If the
constraints are independent, the probability of the Solar System
realizing all of these conditions is given by their product, which is
shown by the dark bell-shaped curve. This joint probability
distribution has an expectation value and variance such that $\langle
N \rangle \approx 4300 \pm 2800$, consistent with the previously
quoted range $N = 10^3 - 10^4$.

The distributions shown in Figure \ref{fig:joint} only place limits on
the membership size $N$ of the putative birth cluster.  Additional
requirements are necessary for successful supernova enrichment, e.g.,
the timing of the supernova and the Solar System location within the
cluster. As a result, the constraints represented by Figure
\ref{fig:joint} are necessary but not sufficient. Note that the
constraints from supernova enrichment can be mitigated, or perhaps
eliminated, if the short-lived radio isotopes are produced by internal
irradiation and distributed supernovae. However, the requirement of a
close encounter to explain Sedna implies almost the same constraint on
the cluster size $N$ as the requirement of a 25 $M_\odot$ progenitor
(see Figure \ref{fig:joint}). As a result, the need for a cluster with
$N = 10^3 - 10^4$ remains.

A wide range of previous studies --- often using different properties
of the Solar System to provide constraints --- have considered the
birth environment of the Sun. In spite of this diversity, many of these
estimates are roughly consistent with the description given above: A
number of authors have highlighted the need for the Sun to form within
some type of cluster in order for supernova enrichment to take place
(Cameron \& Truran 1977, Vanhala \& Boss 2002, Tachibana et al. 2006,
Looney et al.  2006, Megeath et al.  2008). The need for both
supernova enrichment and limited planetary scattering implies a solar
birth cluster with $N \approx 2000 \pm 1000$ (Adams \& Laughlin 2001).
On the other hand, a close encounter with another star in the birth
cluster may be required to explain the observed orbital elements of
Sedna, and perhaps the Kuiper Belt (Brasser et al. 2006, Kenyon \&
Bromley 2004, Morbidelli \& Levison 2004), which suggests that $N
\approx 10^3 - 10^4$ (see also Malmberg et al. 2007).  A similar study
suggests that $N$ = 500 -- 3000, with cluster radius $R$ = 1 -- 3 pc
(Portegies Zwart 2009). If planet scattering is not considered, the expected 
value of cluster membership size $N$ increases. If a large dense
cluster is invoked, so that a large fraction of stars reside near the
center, the cluster membership estimate increases to $N \approx 3
\times 10^5$, with an expected supernova progenitor mass $M_\ast = 75
M_\odot$ (Williams \& Gaidos 2007).  Even more interactive
environments, analogous to the star forming region in the Eagle
Nebula, have also been suggested (Hester et al. 2004). The radiation
fields provided by young clusters are potentially disruptive (Armitage
2000), but the early solar nebula can survive in clusters with $N \le
10^4$ if it spends enough time in the outer regions (Scally \& Clarke
2001, Mann \& Williams 2009).  Finally, chemical considerations
suggest that the Sun formed in the presence of strong FUV radiation
fields, where rough estimates indicate a birth cluster with $N \sim
4000$ (Lee et al. 2008). Although uncertainties remain, these studies
thus suggest that the membership size of the solar birth cluster
should fall in the range $N \approx 10^3 - 10^4$. 

\subsection{Implications for Star and Planet Formation} 

This review of the possible birth environments for the Sun provides an
important consistency check on our current paradigms of star formation
and planet formation. The above considerations suggest that our Solar
System is likely to have formed within a moderately large cluster with
$N = 10^3 - 10^4$, and that the early solar nebula could have been
enriched through both an external supernova and internal irradiation.
The necessary cluster systems are relatively common, with the
Trapezium Cluster in Orion being the closest analog. On a related
note, each of the individual properties of our Solar System that are
affected by the birth environment can be realized with reasonably high
probability (Table II). We thus conclude that the required formation
environment of the Sun is neither rare nor unusual. On the other hand,
the odds of a solar system realizing a particular combination of a
large number of requirements is relatively low.  A quantitative
assessment of the {\it a priori} odds of realizing all of these
properties of our Solar System is difficult to determine, and should
be the subject of further work.

Turning the problem around, this discussion of the solar birth
environment informs our understanding of star and planet formation: We
find that the effects of the birth cluster are neither negligible nor
dominant. The cluster environment can readily sculpt the early solar
nebula, and hence other circumstellar disks, through truncation by
passing stars and especially through evaporation. The resulting
planetary systems can be shaped further by their environment, for
example by changing orbital elements, primarily for companions with
large semimajor axis. Circumstellar disks can acquire significant
quantities of radioactive isotopes from nearby massive stars, and
these nuclei affect their thermal structure. Solar systems can also
gain mass from their environment and readily exchange rocky material
with each other.  The clusters provide ionizing photons, which affect
magnetic coupling of both protostellar cores and circumstellar disks.
On the other hand, major catastrophic events are rare: Disks are
generally not compromised so much that giant planet formation can no
longer (in principle) take place.  Planetary ejection events, driven
by outside influences, are also rare. On average, the cluster
environment thus exerts an intermediate level of influence in
determining solar system properties. Of course, solar systems living
with cluster cores are affected to a much greater extent than those in
the periphery.  On a related note, the effects of clusters on forming
solar systems must be assessed in statistical terms. One important
challenge for the future is thus to determine more accurate
probability distributions for each of the effects discussed herein. We
can then understand in greater detail how the background environment
affects the formation of our Solar System, and others.

\noindent 
$\,$
\bigskip

{\bf Acknowledgments} 

This review benefited from discussions with a large number of
colleagues. I would especially like to thank E. Bergin, M. Duncan,
M. Gounelle, H. Levison, A. Morbidelli, and J. Williams for their
valuable input regarding the manuscript. This work was supported at
the University of Michigan through the Michigan Center for Theoretical
Physics. Portions of this work were carried out at the Isaac Newton
Institute for Mathematical Sciences at Cambridge University.  FCA is
supported by NASA through the Origins of Solar Systems Program (grant
NNX07AP17G), by NSF through the Division of Applied Mathematics (grant
DMS-0806756), and by the Foundational Questions Institute (grant
RFP1-06-1).

\end{document}